\documentclass[final,journal]{IEEEtran}

\IEEEoverridecommandlockouts

\usepackage[T1]{fontenc}
\usepackage{microtype}
\usepackage{amsmath, amssymb, amsfonts}
\usepackage{mathtools}
\usepackage{cite}
\usepackage{tikz,pgfplots}
\usepackage{dblfloatfix}
\usepackage[caption=false]{subfig}
\usepackage{booktabs}
\usepackage[normalem]{ulem}
\usepackage{centernot}
\usepackage{bbm}

\usepackage[applemac]{inputenc}

\usepackage{amsthm}

\newtheorem{assum}{Assumption}
\newtheorem{defn}{Definition}
\newtheorem{lem}{Lemma}
\newtheorem{rem}{Remark}
\newtheorem{prop}{Proposition}
\newtheorem{cor}{Corollary}
\newtheorem{thm}{Theorem}
\pgfplotsset{compat=newest}

\newcommand{\ts}[1]{{\textnormal{#1}}}
\newcommand{\ie}{i.e.,}
\newcommand{\eg}{e.g.}
\newcommand{\mb}{\mathbf}
\newcommand{\mc}{\mathcal}
\newcommand{\mbb}{\mathbb}

\DeclarePairedDelimiter{\ceil}{\lceil}{\rceil}
\DeclarePairedDelimiter{\floor}{\lfloor}{\rfloor}

\DeclarePairedDelimiter{\diagpars}{(}{)}
\newcommand{\diag}{\operatorname{diag}\diagpars}

\DeclarePairedDelimiter{\interiorpars}{(}{)}
\newcommand{\interior}{\operatorname{interior}\interiorpars}

\DeclarePairedDelimiter\abs{\lvert}{\rvert}%
\DeclarePairedDelimiter\norm{\lVert}{\rVert}%

\makeatletter
\let\oldabs\abs
\def\abs{\@ifstar{\oldabs}{\oldabs*}}
\let\oldnorm\norm
\def\norm{\@ifstar{\oldnorm}{\oldnorm*}}
\makeatother

\newif\ifextended
\extendedtrue

\begin{document}

\ifextended
\title{On the Limits of Robust Control Under Adversarial Disturbances---Extended Version}
\fi

\author{Paul Trodden, Jos\'e M. Maestre and Hideaki Ishii%
  \thanks{Paul Trodden is with the School of Electrical and Electronic Engineering, University of Sheffield, Sheffield, UK (email: p.trodden@sheffield.ac.uk)}
  \thanks{Jos\'e M. Maestre is with the Department of Systems and Automation Engineering, University of Seville, Seville, Spain (Email: pepemaestre@us.es)}
  \thanks{Hideaki Ishii is with the Department of Information Physics and Computing, University of Tokyo, Japan (Email: hideaki\_ishii@ipc.i.u-tokyo.ac.jp)}
  \thanks{J. M. Maestre acknowledges support from the JSPS fellowship PE16048 and the MINECO-Spain project C3PO (ref: DPI2017-86918-R).}
}

\maketitle

\begin{abstract}
  This paper addresses a fundamental and important question in
  control: under what conditions does there fail to exist a robust
  control policy that keeps the state of a constrained linear system
  within a target set, despite bounded disturbances? This question has
  practical implications for actuator and sensor specification,
  feasibility analysis for reference tracking, and the design of
  adversarial attacks in cyber-physical systems. While prior research
  has predominantly focused on using optimization to compute
  control-invariant sets to ensure feasible operation, our work
  complements these approaches by characterizing explicit sufficient
  conditions under which robust control is fundamentally
  infeasible. Specifically, we derive novel closed-form, algebraic
  expressions that relate the size of a disturbance set---modelled as
  a scaled version of a basic shape---to the system's spectral
  properties and the geometry of the constraint sets.
\end{abstract}

\section{Introduction}%
\label{sec:Introduction}%

The existence of a robust control policy that maintains system states
within a target set despite disturbances is a fundamental issue in
control. The question of existence is inextricably tied to the notion
of robust control invariance~\cite{Bertsekas72}, a core principle,
important in its own right, underpinning popular techniques such as
Model Predictive Control (MPC)~\cite{RM_mpc_book_2} and Control
Barrier Functions (CBFs)~\cite{AXG+17}. Robust control invariant sets
are state space regions where systems can operate safely under
uncertainty. The existence, characterization and computation of these
sets have been addressed in different contexts by a number of
foundational works~\cite{Bertsekas72,GT91,KG98,Blanchini99,AR08},
spurring a rich body of theoretical knowledge and practical
contributions.

This paper is not concerned with the characterization or computation
of robust control invariant sets, but rather their existence under
different scalings of a bounded disturbance. The impact of disturbance
size on reachability and invariance is well known: given a disturbance
set, it is possible to compute invariant sets that capture the
infinite-time evolution of a system's states or outputs, and it is
understood both in a general sense and in specific instances how these
sets grow, diminish, and even become empty with varying disturbance
set size~\cite{KG98}. We focus on this phenomenon and aim to identify
a disturbance set size sufficient to make robust invariance within a
target set impossible. This inverse problem is not well studied, but,
as several authors point out, is of fundamental relevance to the
design and specification of actuators and sensors~\cite{SSC17},
reference signals and operational limits~\cite{MBZ23}, physical and
controller interconnections in distributed systems~\cite{SMT+23}, and
cybersecurity~\cite{TMI20a}. Indeed, recent work in adversarial
attacks has shown that an attacker can design arbitrarily small, yet
strategically chosen, disturbance sequences that force a control
scheme to violate its state constraints~\cite{DPB+19,ACM22}. This
observation motivates our study of closed-form sufficient conditions
under which no robust policy can exist in the presence of such
adversarial perturbations

The foundational work of De Santis~\cite{DeSantis97} considered a
linear system with parametric uncertainty and bounded additive
disturbances, and developed necessary and sufficient conditions on the
maximal uncertainty set in order that the system state may be robustly
maintained within a specified constraint set; explicit descriptions of
the maximal disturbance sets are given in the case that the
uncertainty sets are polyhedral. In \cite{SSC17}, a stable linear
autonomous system subject to state constraints and additive
disturbances selected from a scaled, bounded disturbance set is
considered. Lower and upper bounds on the critical value of the
disturbance set scaling factor---the smallest value such that
infinite-time robust constraint satisfaction is not possible---are
characterized and computed via optimization-based algorithms. The
authors of~\cite{MBZ23} consider the problem of determining a
polytopic disturbance set such that the corresponding set of reachable
outputs of a stable linear system covers a desired target
polytope. The design of the disturbance set is formulated as a bilevel
optimization problem;~\cite{MBZ24} extends the approach, using notions
of implicit set invariance, to result in a more tractable formulation.

This paper differs from~\cite{DeSantis97,SSC17,MBZ23} in considering
non-autonomous systems rather than autonomous ones. This means we must
tackle the question of existence of a robust control policy in order
to conclude whether a given state constraint set is infinite-time
admissible in response to a disturbance set; the study shifts from
positively invariant sets to control invariant ones. In this regard, a
relevant contribution is~\cite{YRC+20}, which studies the sensitivity
of control invariant sets to parameters, including disturbance
limits. The results are, however, computational rather than algebraic,
a feature of the current paper that distinguishes it
from~\cite{DeSantis97,SSC17,MBZ23} too. We consider a linear
discrete-time system subject to additive disturbances and input
constraints, where it is desired to maintain the state within a target
set.  We exploit the properties of the support function of a convex
set together with the spectral properties of the system's state
transition matrix to develop sufficient conditions for non-existence
of a robust control invariant set and, in turn, robust control
policy. The obtained conditions yield clear and novel insight into how
the non-existence of a robust control policy depends on the system
eigenvalues and the shapes and sizes of the state, input and
disturbance sets. Uniquely, the derived conditions are closed-form and
algebraic, requiring no numerical optimization aside from a few
support function evaluations. This permits us to assume a more general
setting, in which state, input and disturbance sets are merely convex
and compact rather than polyhedral or polytopic.

A preliminary version of this paper appeared in~\cite{TMI20a}, wherein
a linear system subject to input disturbances was considered. The
current paper expands and improves this work in several directions:
firstly, the more general setting of additive state disturbances drawn
from an independent disturbance set is considered. Secondly, the
analysis goes beyond considering only the dominant (and assumed real)
eigenvalue of the open-loop system; the current paper considers a
general linear system with any arrangement and multiciplicity of
eigenvalues, and develops sufficient conditions for non-existence of a
robust control policy that explicitly depend on these
eigenvalues. These conditions reveal distinct behaviours between
stable and unstable systems: for unstable eigenvalues---whether real
positive, real negative, or complex---the critical disturbance size is
largely dictated by the eigenvalue magnitude, while in the stable
case, the geometry of the constraint sets plays a more significant
role, with additional distinctions arising from the sign or complexity
of the eigenvalues.

Finally, we would like to emphasize the novel aspects of this work
from a broader perspective of cybersecurity for control systems. This
area has gained much research interest in recent years (see
\eg~\cite{FT21,IZ22}). In this paper, we consider adversarial attacks
in the form of false-data injections on control inputs. Prior work
addresses zero-dynamics~\cite{TSS15a,PLS19,KI24} and replay
attacks~\cite{MS09,ZM14}. Such studies typically analyse the effects
of the attacks on control systems for the specific attack signals. Our
problem is quite different in that we do not work with given attack
signals but rather study the existence of potential attacks. More
specifically, an attacker can modify the control input arbitrarily but
with a certain bound for causing faulty behaviours in the system. We
derive conditions under which even if the system operator detects the
attacks, there is no way to keep the system state in a safe region
through the control. Furthermore, as we will see, our approach can
handle Denial-of-Service type attacks as well; such attacks can
disable the communication between plant and controller and
compromise stability~\cite{PT15,LSC15,FCI22}.

The paper is organized as follows. Section~\ref{sec:2} gives the
problem statement. In Section~\ref{sec:4}, we recall some established
results on robust constraint admissible and control invariant sets,
and develop some new ones that facilitate our developments. The main
results of the paper are presented in Section~\ref{sec:5}, and
illustrated in Section~\ref{sec:6}, which also shows how the results
may be exploited to design adversarial attack strategies. Conclusions
and directions for future work are presented in
Section~\ref{sec:8}. 

\emph{Notation:} The sets of non-negative and positive reals are
denoted, respectively, $\mbb{R}_{0+}$ and $\mbb{R}_+$. The sets of
natural numbers including and excluding zero are $\mathbb{N}$ and
$\mathbb{N}_+$ respectively. For $a, b \in \mbb{R}^n$, $a \leq b$
applies element by element. For $X, Y \subset \mbb{R}^n$, the
Minkowski sum is $X \oplus Y \coloneqq \{ x + y: x \in X, y \in Y \}$;
for $Y \subset X$, the Minkowski difference is
$X \ominus Y \coloneqq \{ x \in \mbb{R}^n: Y + x \subset X \}$. For
$X \subset \mbb{R}^n$ and $a \in \mbb{R}^n$, $X \oplus a$ means
$X \oplus \{a\}$. $AX$ denotes the image of a set
$X \subset \mbb{R}^n$ under the linear mapping
$A : \mbb{R}^n \to \mbb{R}^p$, and is given by $\{ Ax : x \in
X\}$. The set $-X \coloneqq \{-x: x \in X\}$ is the image of $X$ under
reflection in the origin. The support function of a non-empty set
$X \subset \mbb{R}^n$ is
$h_X(x) \coloneqq \sup \{x^\top z : z \in X \}$. The indicator
function of a set $X$ is $\mathbbm{1}_X$, defined such that
$\mathbbm{1}_X(x) = 0$ if $x\in X$ and $+\infty$ otherwise. A C-set is
a convex and compact (closed and bounded) set containing the origin; a
PC-set is a C-set with the origin in its interior.

\section{Problem statement}%
\label{sec:2}

This section formalizes the problem considered in the paper. We
describe the system dynamics, the disturbance model, and the
constraints, and define the notion of admissible control laws. The
goal is to determine conditions under which no such law can robustly
enforce the constraints. The section concludes by relating the problem
to concepts from robust control invariance and adversarial attack
design.

We consider a discrete-time, linear time-invariant system,
\begin{subequations}\label{eq:dyn}
\begin{align}
  x_{k+1} &= Ax_k + Bu_k + w_k,\\
  (x_k,u_k,w_k) &\in X \times U \times W,
  \end{align}
\end{subequations}
where $x_k \in \mbb{R}^n$, $u_k \in \mbb{R}^m$, $w_k \in \mbb{R}^n$
are, respectively, the state, control input, and disturbance at time
$k\in\mbb{N}$. The state is required to remain within a constraint set
$X \subset \mbb{R}^n$, while the control input must be selected from a
set $U \subset \mbb{R}^m$. The disturbance is drawn from a set
$W \subset \mbb{R}^n$.

\begin{assum}\label{assump:basic}
  The pair $(A,B)$ is reachable; $X$, $U$ and $W$ are PC-sets; $x_k$ is known, but $w_k$ is unknown, at each $k\in\mbb{N}$.
  \end{assum}

  The information pattern considered is therefore the standard one
  from robust control, where the control input $u_k$ may be selected
  using knowledge of $x_k$ but without knowledge of $w_k$.
  \begin{defn}
    An \emph{admissible control law} is a sequence of functions
    $\{\mu_0,\mu_1,\dots\}$ such that
    \begin{equation}
      (\forall k \in \mbb{N})\quad \mu_{k} \colon \mbb{R}^n \to U.
      \end{equation}
  \end{defn}
  The use of an admissible control law of the form
  \begin{equation}
    u_k = \mu_k(x_k)
  \end{equation}
  therefore satisfies the input constraint
  $u_k\in U$ by definition.
  
  In this paper we are interested in characterizing the situation
  where \emph{there does not exist an admissible control law that is
    able to maintain the state $x_k$ in $X$ for all time, despite the
    disturbances}.   The
  following classical definitions help form a more precise statement
  of this aim.
  
  \begin{defn}[Bertsekas~\cite{Bertsekas72}]
  A set $Y \subset \mbb{R}^n$ is:
  \begin{enumerate}
  \item \emph{Infinitely reachable} if there exists an admissible
    control law such that
    \begin{equation}
      (\exists x_0 \in Y)(\forall k\geq 0)(\forall w_k \in W) \quad x_k \in Y.
      \end{equation}
    \item \emph{Strongly reachable} or \emph{robust control invariant} (RCI) if there exists an admissible control law such that
      \begin{equation}\label{eq:prop}
      (\forall x_0 \in Y)(\forall k\geq 0)(\forall w_k \in W) \quad x_k \in Y.
      \end{equation}
       \end{enumerate}
  \end{defn}
  The two notions are linked by the following result. 
  \begin{lem}[Bertsekas~\cite{Bertsekas72}]\label{lem:1}
        The set $X$ is infinitely reachable if, and only if, it contains a robust control invariant set. 
  \end{lem}

  The specific setting considered in the current paper is where the disturbance set is a scaled version of a basic shape set $\bar{W}$:
  \begin{equation}
    W = \alpha \bar{W}, \quad \text{with} \ \alpha > 0.
  \end{equation}
  Our specific aim is then to produce sufficient conditions on the
  parameter $\alpha$ such that $X$ \emph{does not} contain a robust
  control invariant set. By Lemma~\ref{lem:1}, this is equivalent to
  identifying suitable $\alpha$ such that it is \emph{not possible} to
  keep the state robustly in $X$, because there does not exist an
  admissible control law that achieves property~\eqref{eq:prop}.

  From the cyber security perspective, the interpretation of the
  problem is as follows: suppose that a malicious attacker can inject
  false data into the control signal through the form of the
  disturbance $w$ constrained in $\alpha \bar{W}$ and aims at driving
  the state out of the state set $X$. With the solution of our
  problem, the attacker will have to monitor the state evolution to
  find the right moment to start injecting the designed attack signal
  to achieve his goal. In Section~\ref{sec:attacks}, we will address
  the question of how to design the attack signal.

\section{Robust constraint-admissible and control invariant sets }
\label{sec:4}

This section introduces the theoretical tools used to analyze the
existence of robust control invariant sets. We begin by recalling
classical definitions and properties of robust constraint-admissible
sets and their computation. Then, we develop new results that offer
analytical insight into how the system dynamics and disturbance
scaling affect the non-existence of robust policies.

\subsection{Preliminaries}

The $k$-step robust constraint-admissible set is the set of all states
that can be kept within $X$ for at least $k$ time steps, for any
disturbance, respecting the input constraints:
\begin{equation}
(\forall k \in \mbb{N})\ C_k \coloneqq  \bigl\{ x_0 : (\forall \mb{w}_k \in \mc{W}_k)(\exists \mb{u}_k \in \mc{U}_k) \ \mb{x}_k \in \mc{X}_k \bigr\}
\end{equation}
where $\mb{u}_k$ (respectively $\mb{w}_k$) is the sequence of $k$
controls $\{u_0,u_1,\dots,u_{k-1}\}$ (disturbances
$\{w_0,w_1,\dots,w_{k-1}\}$), the set
$\mc{U}_k \coloneqq {U} \times \dots \times {U}$, with a similar
definition relating $ \mc{W}_k$ and ${W}$. The corresponding sequence
$\mb{x}_k = \{x_0,x_1,\dots,x_k\}$ is obtained by, starting from
$x_0$, applying the input sequence $\mb{u}_k$ and disturbance sequence
$\mb{w}_k$ to~\eqref{eq:dyn}. The definition requires
$\mb{x}_k \in \mc{X}_k \coloneqq X \times \dots \times X$.

We recall some basic facts about $C_k$~\cite{Bertsekas72,Blanchini94,Kerrigan00}:
\begin{lem}
  Suppose $U$ is a PC-set and $ W$ is a C-set. Then (i) $C_0 = X$;
  (ii) if $X$ is compact [convex], then each $C_k$ is closed [convex];
  (iii) $C_{k+1} \subseteq C_k$; (iv)
  $C_k = \bigcap_{\ell=0}^k C_\ell$; (v)
  $C_{\infty} \coloneqq \lim_{k \to \infty} C_k =
  \bigcap_{\ell=0}^{\infty} C_\ell$; (vi) if
  $0 \in \interior{C_\infty}$ then every $C_k$, $k\in \mbb{N}$, is a
  PC-set; (vii) if $0 \in \interior{C_{\infty}}$, then ${C}_\infty$ is
  a robust control invariant set for the system~\eqref{eq:dyn}; (viii)
  $C_{\infty} $, if non-empty, is maximal in the sense that it
  contains all other robust control invariant sets for the
  system~\eqref{eq:dyn}.
\end{lem}

The following recursion characterizes the set $C_k$ for each $k$ starting from $C_0 = X$:
\begin{equation}\label{eq:Crecursion}
  (\forall k \in \mbb{N}) \quad C_{k+1} = Q(C_{k}) \cap X,
  \ \text{with} \ C_0 = X,
\end{equation}
where $Q(\cdot)$ is the backwards reachability operation:
\begin{equation} Q(Y) \coloneqq \{ x : (\exists u \in U) \ \{Ax + Bu\} \oplus  W \subseteq Y \}.
\end{equation}
For the linear time invariant
system~\eqref{eq:dyn}, therefore,
\begin{equation}\label{eq:9}
(\forall k \in \mbb{N}) \quad C_{k+1} = (A)^{-1} \bigl([C_{k} \ominus W] \oplus (-BU) \bigr) \cap X,
\end{equation}
where $(A)^{-1}(\cdot)$ denotes the pre-image of the linear
transformation $A(\cdot)$, and exists regardless of whether $A$ is
invertible; for shorthand we will write $A^{-k} Y$ to denote $(A^k)^{-1}(Y)$.

The following lemma links the aim of the paper to the characterization of $C_k$.

\begin{lem}
  If, for some $k^\star > 0$, $C_{k^*} = \emptyset$ then
\begin{enumerate}
\item $C_k = \emptyset$ for all $k \geq k^\star$;
\item $C_\infty = \emptyset$;
  \item $X$ does not contain a robust control invariant set.
  \end{enumerate}
\end{lem}

An equivalent statement of our aim is therefore to determine suitable
$\alpha$ such that $C_k$ is empty for some finite $k$.

\subsection{Some new results}
Inspection of~\eqref{eq:Crecursion}--\eqref{eq:9} reveals that the determination of suitable $\alpha$ such that some $C_{k^\star}$ is empty requires the analysis of the
sequence of sets $\{C_k\}$. Yet the dynamics of these sets are characterized by a sequence of Minkowski addition and subtractions, set intersections and preimages, and are not
readily amenable to analysis. The following result, which
appears to be new, therefore gives insight into how $C_k$ (and hence
$C_\infty$) may be characterized in terms of sets with simpler
dynamics that, as we will show, are more amenable to analysis. A proof is in Appendix~\ref{proof:Tdef}.

\begin{prop}\label{prop:Tdef}
  The set $C_{k}$ is bounded as
  \begin{align}
    (\forall k \in \mbb{N})\quad C_k &\subseteq \bigcap_{\ell=0}^{k} A^{-\ell} T_\ell \label{eq:Cinclusion}
          \intertext{where}
          \begin{split}
               (\forall k \in \mbb{N})\quad T_{k+1} &\coloneqq (T_k \ominus A^k W) \oplus A^k(-BU) \\ 
    \text{with} \ T_0 &= X. \end{split}
  \end{align}
  
\end{prop}

\begin{rem}\label{rem:tight}
  
  A special case of this result was reported by~\cite{SSC17}, wherein
  an autonomous system $x_{k+1} = Ax_k + w_k$ subject to a
  disturbance from a scaled set $\alpha W$ is considered. In that setting,
  \ie~without a control input, what we refer to here as $C_\infty$ is the maximal robust
  \emph{positively} invariant set. The authors determine conditions on
  the scaling constant $\alpha$ under which $C_\infty$
  exists and, moreover, develop the following
  relation~\eqref{eq:Schulze}, which we show now to be a
  corollary of Proposition~\ref{prop:Tdef}.
   \begin{cor} If ${U} = \{0\}$ then ${T}_k = {X} \ominus {R}_k$ and
      \begin{equation}\label{eq:Schulze}
      {C}_k = \bigcap_{\ell=0}^k (A^\ell)^{-1}( {X} \ominus {R}_k) \ \text{where} \ R_k \coloneqq \bigoplus_{j=0}^k A^j W.
\end{equation}
\end{cor}
Note that in~\eqref{eq:Schulze} the relation for $C_k$ holds with
equality, and not just the inclusion depicted in
Proposition~\ref{prop:Tdef}. The reason for the weakening of the
equality to mere inclusion when $U \supset \{ 0 \}$ is the behaviour
of the Minkowski sum under intersection: for sets $A$, $B$ and $C$,
$(A \cap B) \oplus C \subseteq (A\oplus C) \cap (B \oplus C)$ and not
$(A \cap B) \oplus C = (A\oplus C) \cap (B \oplus C)$. The latter
equality does hold if the \emph{union} of convex sets $A$ and $B$ is
convex, but this is generally not the case. This, together with the
fact that we consider a non-autonomous system and not a stable
autonomous one, means that the methods and results of~\cite{SSC17}
cannot be applied to the problem considered in this paper.
\end{rem}

We conclude the section by establishing sufficient conditions for
emptiness of the set $C_k$ for some $k > 0$ and subsequently
$C_\infty$. The results are central to the subsequent
developments. Proofs of Propositions~\ref{prop:Cexist}
and~\ref{prop:Texist} are provided in Appendices~\ref{proof:Cexist}
and~\ref{proof:Texist}, respectively.

\begin{prop} 
        If, for some $k^\star > 0$, $T_{k^\star} = \emptyset$ then
        $C_{k} = \emptyset$ for all $k \geq k^\star$.
      \label{prop:Cexist}
      \end{prop}

\begin{prop} 
 If, for some $k^\star > 0$, $S_{k^\star} = \emptyset$, where
    \begin{equation}\label{eq:Scond}
     (\forall k \in \mbb{N})\quad S_k \coloneqq X \oplus \left[\bigoplus_{\ell=0}^{k-2} A^\ell (-BU)\right] \ominus \left[\bigoplus_{\ell=0}^{k-1} A^\ell W \right],
      \end{equation}
      then $T_{k^\star} = \emptyset$.
      \label{prop:Texist}
      \end{prop}

      Verifying emptiness of $S_k$ is easily recast as a problem of
      checking some inequalities, via the well known necessary and
      sufficient conditions based on support functions:

      \begin{lem}\label{lem:Ssupport}
        The set $S_k$ is empty if and only if there exists a
        $z \in \mbb{R}^n$ such that
        \begin{equation}\label{eq:supc}
          h_{S_k}(z) < 0 \ \text{and} \ h_{S_k}(-z) < 0.
        \end{equation}
        \end{lem}

        \begin{rem}
          Note that if the sets $X$, $U$ and $\bar{W}$ are
          symmetrical, then condition~\eqref{eq:supc} becomes
          $h_{S_k}(z) < 0$. Consequently, all of the remaining results
          in the paper, which rely on evaluating support functions in
          the directions $\pm z$, may be simplified to checking in
          either the direction $+z$ or the direction $-z$.
          \end{rem}

          \section{For which values of $\alpha$ is $C_\infty$ empty?}
          \label{sec:5}

      In this section, we specialize the general results of
      Section~\ref{sec:4} to the specific case of a disturbance set
      defined by a scaling parameter $\alpha$. Our aim is to derive
      conditions on $\alpha$ under which the robust control invariant
      set is guaranteed to be empty. We do this by first bounding the
      admissible set sequence and then developing closed-form,
      algebraic conditions that relate $\alpha$ to the system's
      eigenstructure and the geometry of the constraint sets. The main
      result is a family of sufficient conditions for infeasibility
      that can be efficiently evaluated and interpreted in terms of
      system behavior.

We now recast some of the results from the previous section in the
particular setting of the paper, and develop conditions under which
$C_\infty$ is empty. Considering that $W = \alpha\bar{W}$, let
\begin{align}
  \begin{split}
  (\forall k \in \mbb{N})\ C^{\alpha}_{k+1} &\coloneqq A^{-1}([C_{k}^{\alpha} \ominus (\alpha \bar{W})] \oplus  (-B{U})) \cap X\\
  \text{with} \ C^{\alpha}_0 & = {X},\label{eq:Calpha}
  \end{split}
                                      \intertext{and}
                                      \begin{split}
 (\forall k \in \mbb{N})\ {T}^{\alpha}_{k+1} &\coloneqq ({T}^{\alpha}_k \ominus  A^k (\alpha\bar{W}) ) \oplus (A^k(-B{U}))\\
 \text{with} \ {T}^{\alpha}_0 &= {X},
 \end{split}\label{eq:Titer}
  \end{align}
  where the sets are super-indexed by $\alpha$ to denote their
  dependency on the scaling factor. The connection between the two is, following Proposition~\ref{prop:Tdef},
  \begin{equation}
    (\forall k \in \mbb{N}) \ C_{k}^{\alpha} \subseteq \bigcap_{\ell=0}^k A^{-\ell} T^{\alpha}_k.
    \end{equation}
    In a similar way, the set $S_k$ in Proposition~\ref{prop:Texist}
    may be specialized to the setting and denoted $S_k^{\alpha}$:
\begin{equation}\label{eq:Scond2}
     (\forall k \in \mbb{N}) \ S^\alpha_k \coloneqq X \oplus \left[\bigoplus_{\ell=0}^{k-2} A^\ell (-BU)\right] \ominus \left[\bigoplus_{\ell=0}^{k-1} A^\ell (\alpha\bar{W}) \right].
      \end{equation}

      Our goal is to characterize, for each $k \in \mbb{N}$, the
      scaling factor $\alpha$ that results in $C_{k}^{\alpha}$ being
      empty. The smallest such value of $\alpha$, for a given $k$,
    \begin{equation}
      \alpha^*_{k} \coloneqq \inf \bigl\{ \alpha \geq 0 : C_{k}^{\alpha} = \emptyset \bigr\},
    \end{equation}
    determines the smallest scaling factor such
    that the state cannot be kept robustly within $X$ for more
    than $k$ steps.
    
In Theorem~\ref{thm:main}, we establish an upper bound on
$\alpha^*_{k}$ for any $k$. We achieve this by characterizing, for
each $k \in \mbb{N}$, a value $\bar{\alpha}^+_k$ that renders
$S^{\alpha}_{k}$ empty for all $\alpha > \bar{\alpha}^+_{k}$. By
Proposition~\ref{prop:Texist} and Corollary~\ref{prop:Cexist}, any
$\alpha > \bar{\alpha}^+_{k} \geq \alpha^*_{k}$ then ensures that
$C^{\alpha}_{k}$ is empty. We find that these bounds depend on the
relative sizes and shapes of the sets $X$, $U$ and $\bar{W}$, as well
as the stability or instability (via the spectral properties)
of the open-loop system.

One way of finding
the critical scaling factor is to compute, for different combinations of
$\alpha$ and $k$, the support function of $S^{\alpha}_{k}$ (or, indeed, $C_k^\alpha$) in a number
of directions $\pm z_1,\dots,\pm z_p \in \mbb{R}^n$ and check for negativity. Each support
function evaluation is a linear programming problem. Such an approach,
however, scales poorly with the state dimension and, besides, merely
provides a numerical result specific to a considered system rather
than a more general, analytical result. In this paper, we pursue the
latter. The development, inspired by~\cite{TM04}, is enabled by
considering the real Jordan form of the system~\eqref{eq:dyn}:
    \begin{lem}
      For $A \in \mbb{R}^{n \times n}$ in~\eqref{eq:dyn} there exists a real-valued nonsingular matrix $\Phi$ and a real-valued matrix $J$ such that $\Phi^{-1} A^\top \Phi = J = \diag{J_1,\dots,J_N}$, where $J_i$ is the real Jordan block given by
\begin{equation}
        J_i = 
          \begin{bmatrix} D_i & \mb{1}_i & & \\ & D_i & \mb{1}_i & \\ & & \ddots & \mb{1}_i \\ & & & D_i \end{bmatrix},
      \end{equation}
      with
      \begin{align}
        D_i &\coloneqq \begin{cases}
          \lambda_i, & \text{if $\lambda_i$ is real}, \\
          \rho_i R(\theta_i),
          & \text{if $\lambda_i = \rho_i\left( \cos \theta_i \pm \jmath \sin \theta_i\right) $,}
        \end{cases}\label{eq:D}\\
        R(\theta_i) &\coloneqq \begin{bmatrix*}[r] \cos \theta_i & -\sin \theta_i \\ \sin\theta_i & \cos\theta_i \end{bmatrix*},\label{eq:R}\\
        \mb{1}_i &\coloneqq \begin{cases}
          1, & \text{if $\lambda_i$ is real}, \\
          I_{2 \times 2},
          & \text{if $\lambda_i = \rho_i\left( \cos \theta_i \pm \jmath \sin \theta_i\right)$.}
        \end{cases}
        \end{align}
    \end{lem}

    In the sequel, we develop sufficient conditions on the scaling
    parameter $\alpha$ for emptiness of $C_{\infty}$ by exploiting the
    structure revealed by the Jordan decomposition of $A^\top$, the
    dual of $A$. The main idea is that, by evaluating the
    condition~\eqref{eq:supc} with $z$ chosen from the columns of the
    modal matrix $\Phi$, closed-form algebraic conditions emerge in
    place of the iterative condition~\eqref{eq:Scond}. We first
    consider the case where an eigenvalue $\lambda_i$ is real and
    positive, and subsequently, in Section~\ref{sec:general}, extend
    the results to cover negative--real and complex eigenvalues.

    \subsection{The eigenvalue $\lambda_i$ is real and positive}\label{sec:eig}

    Consider the $i$th Jordan block $J_i$, of dimension $n_i$, and suppose the
    associated eigenvalue $\lambda_i$ is real and positive. Let
    \begin{equation}
      \Phi_i =
      \begin{bmatrix}
        \varphi_{i,1}&
        \varphi_{i,2}&
        \dots&
        \varphi_{i,n_i}
        \end{bmatrix}
      \end{equation}
      be the matrix containing the $n_i$ columns of the modal matrix
      $\Phi$ corresponding to the block $J_i$. The first column of
      $\Phi_i$ is an eigenvector of $A^\top$ (therefore a left
      eigenvector of $A$) satisfying
      \begin{equation}\label{eq:23}
         A^\top \varphi_{i,1} = \lambda_{i} \varphi_{i,1},
        \end{equation}
        while the remaining $(n_i-1)$ columns of $\Phi_i$ are generalized eigenvectors, satisfying
        \begin{equation}\label{eq:24}
         (j=2,\dots,n_i) \ A^\top \varphi_{i,j} =  \lambda_i \varphi_{i,j} + \varphi_{i,j-1}.
          \end{equation}
          The recursive application of~\eqref{eq:23} and~\eqref{eq:24} yields, for $\ell \geq 0$,
          \begin{equation}\label{eq:AD}
            (A^\ell)^\top \varphi_{i,j} = \begin{cases} \lambda_i^\ell \varphi_{i,1}, & j = 1,\\
              \sum_{p = 0}^{j-1} \binom{\ell}{p} \lambda_i^{\ell-p} \varphi_{i,j-p}, & j = 2,\dots,n_i,\end{cases}
            \end{equation}
            where $\binom{\ell}{p}$ is the binomial coefficient. This
            provides simple algebraic relations that describe how the
            vectors $\varphi_{i,j}$ evolve under repeated application
            of the dual dynamics
            $\varphi_{i,j}^+ = A^\top \varphi_{i,j}$.

This is important because testing the condition~\eqref{eq:supc} for
the set $S_k^{\alpha}$ defined in Proposition~\ref{prop:Texist}
requires evaluating support functions of sets of the form $A^\ell Y$. But it
is well known that
\begin{equation}
  (\forall z \in \mbb{R}^n)\ h_{A^\ell Y}(z) = h_Y((A^\ell)^\top z), 
\end{equation}
so that the support functions of different sets
$A^0 Y, A^1 Y, \dots, A^\ell Y$ in the same direction $z$ can be
expressed as the support function of the same set $Y$ in different
directions $(A^0)^\top z, (A^1)^\top z, \dots, (A^\ell)^\top
z$. Equation~\eqref{eq:AD} indicates that if we choose the direction $z$
in which to evaluate the support as one of the (generalized)
eigenvectors contained in $\Phi_i$, the support functions
$h_{A^0 Y}(z), h_{A^1 Y}(z), \dots, h_{A^\ell Y}(z)$ potentially become simple,
closed-form algebraic functions of $h_Y(z)$. For example, when considering the direction of the eigenvector associated with a real, positive eigenvalue, the support function of the a sey $A^{l}Y$ in the direction $z=\pm\varphi_{i,1}$ when eigenvalue $\lambda_{i}>0$ is
\begin{equation}\label{eq:30}
  h_{A^\ell Y}(\pm \varphi_{i,1}) = h_Y((A^\ell)^\top (\pm \varphi_{i,1})) = \lambda^\ell_i h_Y(\pm \varphi_{i,1}),
\end{equation}
which uses the support function property $h_Y(az) = ah_Y(z)$
for $a \geq 0$~\cite{Schneider93}.

Consequently, our first main result, Theorem~\ref{thm:main}, arises from evaluating the support function
of $S_k^{\alpha}$, and interrogating the condition~\eqref{eq:supc}, in the directions $z = \pm \varphi_{i,1}$ associated with a real positive eigenvalue. Using~\eqref{eq:30}, we have
\ifCLASSOPTIONonecolumn
\begin{equation}\label{eq:31}
(\forall k \in \mbb{N})\ {h}_{S_k^{\alpha}}(\pm \varphi_{i,1}) \leq {h}_X(\pm \varphi_{i,1})  + \sum_{\ell=0}^{k-2} {h}_{A^\ell (-BU)}(\pm \varphi_{i,1}) - \sum_{\ell=0}^{k-1} {h}_{A^\ell (\alpha\bar{W})}(\pm \varphi_{i,1}) ,
\end{equation}
\else
\begin{multline}\label{eq:31}
(\forall k \in \mbb{N})\ {h}_{S_k^{\alpha}}(\pm \varphi_{i,1}) \leq {h}_X(\pm \varphi_{i,1})  \\+ \sum_{\ell=0}^{k-2} {h}_{A^\ell (-BU)}(\pm \varphi_{i,1}) - \sum_{\ell=0}^{k-1} {h}_{A^\ell (\alpha\bar{W})}(\pm \varphi_{i,1}) ,
\end{multline}
\fi
and, thence, by the properties of the eigenvector dynamics and support function, the simple algebraic expression
\ifCLASSOPTIONonecolumn
\begin{equation}\label{eq:halpha1}
(\forall k \in \mbb{N})\ {h}_{S_k^{\alpha}}(\pm \varphi_{i,1}) \leq {h}_X(\pm \varphi_{i,1})  + {h}_{-BU}( \pm \varphi_{i,1}) \sum_{\ell=0}^{k-2} \lambda_i^\ell   - \alpha {h}_{\bar{W}}(\pm \varphi_{i,1}) \sum_{\ell=0}^{k-1}\lambda_i^\ell .
\end{equation}
\else
\begin{multline}\label{eq:halpha1}
(\forall k \in \mbb{N})\ {h}_{S_k^{\alpha}}(\pm \varphi_{i,1}) \leq {h}_X(\pm \varphi_{i,1}) \\ + {h}_{-BU}( \pm \varphi_{i,1}) \sum_{\ell=0}^{k-2} \lambda_i^\ell   - \alpha {h}_{\bar{W}}(\pm \varphi_{i,1}) \sum_{\ell=0}^{k-1}\lambda_i^\ell .
\end{multline}
\fi

Theorem~\ref{thm:main} characterizes a sufficient condition under which the right-hand side of~\eqref{eq:halpha1} is negative, and hence $S_k^\alpha$, $T_k^\alpha$ and $C_k^\alpha$ are empty.
                    
\begin{thm}\label{thm:main}
  Consider the $ith$ Jordan block $J_i$ and suppose that the
  associated eigenvalue $\lambda_i$ is real and positive. If, for some
  $k^\star \in \mbb{N}_+$,
  \begin{equation}\label{eq:alpha111}
    \alpha > \bar{\alpha}_{k^\star}^+(\lambda_i) \coloneqq \max\{ \alpha^+_{k^\star}(+\varphi_{i,1}), \alpha^+_{k^\star}(-\varphi_{i,1})\}
  \end{equation}
  where, for $k\in\mbb{N}_+$ and $z = \pm \varphi_{i,1}$,
  \ifCLASSOPTIONonecolumn
\begin{equation}\label{eq:alpha1}
    \alpha^+_{k}(z) \coloneqq
    \begin{cases}
      \left(\frac{1-\lambda_i}{1-\lambda_i^{k}}\right)\frac{h_{X}(z)}{h_{\bar{W}}(z)} + \left(\frac{1-\lambda_i^{k-1}}{1-\lambda_i^{k}}\right)\frac{h_{-BU}(z)}{h_{\bar{W}}(z)}, & \lambda_i \neq 1, \\
      \left(\frac{1}{k}\right)
      \frac{h_{X}(z)}{h_{\bar{W}}(z)} +
      \left(\frac{k-1}{k}\right)
      \frac{h_{-BU}(z)}{h_{\bar{W}}(z)},&
      \lambda_i = 1,
    \end{cases}
  \end{equation}
  \else
\begin{multline}\label{eq:alpha1}
    \alpha^+_{k}(z) \coloneqq\\
    \begin{cases}
      \left(\frac{1-\lambda_i}{1-\lambda_i^{k}}\right)\frac{h_{X}(z)}{h_{\bar{W}}(z)} + \left(\frac{1-\lambda_i^{k-1}}{1-\lambda_i^{k}}\right)\frac{h_{-BU}(z)}{h_{\bar{W}}(z)}, & \lambda_i \neq 1, \\
      \left(\frac{1}{k}\right)
      \frac{h_{X}(z)}{h_{\bar{W}}(z)} +
      \left(\frac{k-1}{k}\right)
      \frac{h_{-BU}(z)}{h_{\bar{W}}(z)},&
      \lambda_i = 1,
    \end{cases}
  \end{multline}
  \fi%
  then $C^{\alpha}_{k} = \emptyset$ for all $k \geq k^\star$.
\end{thm}

This result, and many that follow, makes extensive use of
            finite geometric series and particularly the closed-form
            expressions for their partial sums. The relevant
            expressions are given in the following lemma.

            \begin{lem}
              The following expressions hold:
              \ifCLASSOPTIONonecolumn
              \begin{align}
 s_{a,b}(x) &\coloneqq \sum_{i=a}^{b} x^i = \begin{cases} \frac{x^a - x^{b+1}}{1 - x}, & x \neq 1, \\
    b-a+1, & x = 1.
  \end{cases}\label{eq:geo1}\\
       s^\prime_{a,b}(x)& \coloneqq         \sum_{i=a}^{b} i x^{i-1} = \begin{cases} \frac{ax^{a-1}+(1-a)x^a + bx^{b+1} - (b+1)x^b}{(1-x)^2}, & x \neq 1, \\
    \frac{(b-a+1)(a+b)}{2}, & x = 1.\label{eq:geo2}
  \end{cases}
              \end{align}
              \else
\begin{align}
 s_{a,b}(x) &\coloneqq \sum_{i=a}^{b} x^i = \begin{cases} \frac{x^a - x^{b+1}}{1 - x}, & x \neq 1, \\
    b-a+1, & x = 1.
  \end{cases}\label{eq:geo1}\\
                \begin{split}
       s^\prime_{a,b}(x)& \coloneqq         \sum_{i=a}^{b} i x^{i-1} \\ &= \begin{cases} \frac{ax^{a-1}+(1-a)x^a + bx^{b+1} - (b+1)x^b}{(1-x)^2}, & x \neq 1, \\
    \frac{(b-a+1)(a+b)}{2}, & x = 1.\label{eq:geo2}
  \end{cases}
  \end{split}
              \end{align}
              \fi

              \end{lem}

              \textit{Proof of Theorem~\ref{thm:main}:}
      To establish the right-hand side of~\eqref{eq:halpha1} to be negative, consider \eqref{eq:Scond2} and the support function of this set evaluated for
      $z = \pm \varphi_{i,1}$:
      \ifCLASSOPTIONonecolumn
\begin{equation}\label{eq:proofh}
(\forall k \in \mbb{N}) \ {h}_{S_k^{\alpha}}(\pm\varphi_{i,1}) \leq {h}_X(\pm\varphi_{i,1}) + \sum_{\ell=0}^{k-2} {h}_{A^\ell (-BU)}(\pm\varphi_{i,1}) - \sum_{\ell=0}^{k-1} {h}_{A^\ell (\alpha\bar{W})}(\pm\varphi_{i,1}) ,
\end{equation}
      \else
\begin{multline}\label{eq:proofh}
(\forall k \in \mbb{N}) \ {h}_{S_k^{\alpha}}(\pm\varphi_{i,1}) \leq {h}_X(\pm\varphi_{i,1}) \\ + \sum_{\ell=0}^{k-2} {h}_{A^\ell (-BU)}(\pm\varphi_{i,1}) - \sum_{\ell=0}^{k-1} {h}_{A^\ell (\alpha\bar{W})}(\pm\varphi_{i,1}) ,
\end{multline}
\fi%
where we have used the support function properties $h_{X\oplus Y}(z) = h_X(z)+h_Y(z)$ and $h_{X \ominus Y}(z) \leq h_X(z) - h_Y(z)$ for C-sets $X$ and $Y$. By definition, for a C-set $Y$ and matrix $M$, $h_{MY}(z) = h_Y(M^\top z)$; additionally, $h_{-Y}(z) = h_{Y}(-z)$ and $h_{aY}(z) = h_Y(az) = ah_Y(z)$ for a scalar $a \geq 0$. Therefore 
\ifCLASSOPTIONonecolumn
\begin{equation}
(\forall k \in \mbb{N}) \ {h}_{S_k^{\alpha}}(z) \leq {h}_X(z) + \sum_{\ell=0}^{k-2} {h}_{-BU}((A^\ell)^\top (z))  - \alpha \sum_{\ell=0}^{k-1} {h}_{\bar{W}}((A^{\ell})^\top (z)),
\end{equation}
\else
\begin{multline}
(\forall k \in \mbb{N}) \ {h}_{S_k^{\alpha}}(z) \leq {h}_X(z) \\+ \sum_{\ell=0}^{k-2} {h}_{-BU}((A^\ell)^\top (z))  - \alpha \sum_{\ell=0}^{k-1} {h}_{\bar{W}}((A^{\ell})^\top (z)),
\end{multline}
\fi%
where $z = +\varphi_{i,1}$ or $-\varphi_{i,1}$. Owing to~\eqref{eq:AD}, the right-hand side of this inequality is
\begin{equation}\label{eq:negat}
{h}_X(z) + {h}_{-BU}( z) \sum_{\ell=0}^{k-2} \lambda_i^\ell   - \alpha {h}_{\bar{W}}(z) \sum_{\ell=0}^{k-1}\lambda_i^\ell  .
  \end{equation}
Given that $h_X(\pm\varphi_{i,1}) > 0$ and $h_{\bar{W}}(\pm\varphi_{i,1}) > 0$ by Assumption~\ref{assump:basic}, that $h_{-BU}(\pm\varphi_{i,1}) \geq 0$ by Assumption~\ref{assump:basic} in conjunction with the fact that $-BU$ is a C-set, and $\lambda_i > 0$, the expression~\eqref{eq:negat} is negative for both $z=+\varphi_{i,1}$ \emph{and} $z=-\varphi_{i,1}$ for all pairs $(\alpha,k)$ that satisfy
  \begin{equation}
    \begin{dcases} \begin{aligned} \alpha &> 0, k \geq 1, \\
        \alpha &> \frac{{h}_{X}(+\varphi_{i,1}) + s_{0,k-2}(\lambda_i) {h}_{-BU}(+\varphi_{i,1})}{s_{0,k-1}(\lambda_i)h_{\bar{W}}(+\varphi_{i,1})}\\
        \alpha &> \frac{{h}_{X}(-\varphi_{i,1}) + s_{0,k-2}(\lambda_i) {h}_{-BU}(-\varphi_{i,1})}{s_{0,k-1}(\lambda_i)h_{\bar{W}}(-\varphi_{i,1})}
      \end{aligned}
      \end{dcases}
             \label{eq:abound}
  \end{equation}
  The right-hand sides of the two final inequalities in~\eqref{eq:abound} are ${\alpha}_{k^\star}^+(+\varphi_{i,1})$ and ${\alpha}_{k^\star}^+(-\varphi_{i,1})$, respectively, defined in~\eqref{eq:alpha1}. Therefore, any pair $(\alpha^\star,k^\star)$ satisfying~\eqref{eq:abound} also satisfies
  \begin{equation}
\alpha^\star > \max\{ {\alpha}_{k^\star}^+(+\varphi_{i,1}), {\alpha}_{k^\star}^+(-\varphi_{i,1}) \}.
    \end{equation}
  It follows that $S_{k^\star}^{\alpha^\star} \subset S_{k^\star}^{\bar{\alpha}_{k^\star}^+(\lambda_i)}$, that $S_{k^\star}^{\alpha^\star}$ is empty, and, in fact, that $S^{\alpha}_{k^\star} = \emptyset$ for all $\alpha > \bar{\alpha}_{k^\star}^+(\lambda_{i})$. In turn, by Corollary~\ref{prop:Cexist} and Proposition~\ref{prop:Texist}, if $\alpha > \bar{\alpha}_{k^\star}^+(\lambda_i) $ then $C_k^\alpha = \emptyset$ for all $k \geq k^\star$.
\hfill\IEEEQED

The bound here provides insight into how the system dynamics affect
the minimum size of the disturbance set sufficient to ensure
non-existence of a robust control law. In particular, the critical
value of $\alpha$ depends on the magnitude of the considered
eigenvalue and the relative sizes of the state, input and disturbance
sets in the direction of the corresponding eigenvector; the examples
in Section~\ref{sec:6} illustrate this. Note that the bounds are well
defined because, by Assumption~\ref{assump:basic},
$\bar{W}$ must be a PC-set and, hence, $h_{\bar{W}}(z) > 0$ for all
$z\in\mbb{R}^n$.

\subsection{Closed-form bounds for non-existence of a robust policy}

It remains to determine the \emph{smallest} $\alpha$ sufficient to
make $C^\alpha_\infty$ empty. The issue is this: although finding an
$\alpha$ sufficient to make $C^\alpha_{k}$, in~\eqref{eq:Calpha},
empty ensures that all $C_{k+1}^\alpha, C_{k+2}^\alpha, \ldots$ are
also empty for the same $\alpha$, finding an $\alpha$ that makes
$S_k^\alpha$ empty does not imply that all
$S_{k+1}^\alpha, S_{k+2}^\alpha, \ldots$ are also empty. This is
because of the conservativism induced by considering $S^\alpha_k$ as a
proxy for $C^\alpha_k$. Nonetheless, if $\alpha$ is sufficient to make
$S_k^\alpha$ empty then indeed the same $\alpha$ guarantees that $C_k$
and all $C_{k+1}^\alpha, C_{k+2}^\alpha, \ldots$ are also empty for
the same $\alpha$. We therefore need to find
\begin{equation}
\inf_{k\in\mbb{N}_+} \bar{\alpha}^+_k(\lambda_i)
  \end{equation}
  as the closest approximation to the smallest $\alpha$ that guarantees $C^\alpha_\infty = \emptyset$:
  \begin{equation}
\alpha_\infty^* \coloneqq \inf\{ \alpha \geq 0 : C^\alpha_\infty = \emptyset\}.
    \end{equation}

For this, it is useful to study the behaviour of the bound defined in Theorem~\ref{thm:main}.

\begin{prop}\label{prop:alphaseq}
  The sequence~$\{\alpha^+_{k}(z)\}_{k \in \mbb{N}_+}$, defined by~\eqref{eq:alpha1} with $z=+\varphi_{i,1}$ or $z=-\varphi_{i,1}$, is positive and:
  \begin{enumerate}
    \item monotonically decreasing if $\lambda_i h_X(z)>h_{-BU}(z) $;
    \item monotonically increasing if $\lambda_i h_X(z)<h_{-BU}(z) $;
      \item constant if $\lambda_i h_X(z)=h_{-BU}(z) $.
    \end{enumerate}
    Consequently, for $z=+\varphi_{i,1}$ or $-\varphi_{i,1}$,
  \begin{equation}
    \inf_{k\in\mbb{N}_+} \alpha^+_k(z) = \begin{cases} {\alpha}_1^+(z), & \lambda_i h_X(z)\leq h_{-BU}(z), \\
      {\alpha}_{\infty}^+(z) , & \lambda_i h_X(z)> h_{-BU}(z).
    \end{cases}
  \end{equation}
  where ${\alpha}_{\infty}^+(z) \coloneqq \lim_{k\to\infty} {\alpha}_{k}^+(z)$.
  \end{prop}

  \begin{IEEEproof}
    Positivity of both ${\alpha}_k^+(+\varphi_{i,1})$ and ${\alpha}_k^+(-\varphi_{i,1})$ follows from $h_X(z), h_{-BU}(z), h_{\bar{W}}(z) > 0$ for all $z\in\mbb{R}^n$ and the series $s_{0,k} > 0$ for all $k \in \mbb{N}_+$, where we use the shorthand notation $s_{0,k}$ for $s_{0,k}(\lambda_i)$. Consider ${\alpha}_k^+(z)$ and $\alpha_{k+1}^+(z)$ for $z=+\varphi_{i,1}$ or $-\varphi_{i,1}$. From~\eqref{eq:abound},
    \ifCLASSOPTIONonecolumn
\begin{equation}\label{eq:diffalpha}
  \bar{\alpha}_{k+1}^+(z) - \bar{\alpha}_k^+(z) =
  \frac{(s_{0,k-1}-s_{0,k}) h_X(z) + (s_{0,k-1}^2 - s_{0,k-2}s_{0,k} ) h_{-BU}(z)}{s_{0,k}s_{0,k-1} h_{\bar{W}}(z)}.
\end{equation}
    \else
\begin{multline}\label{eq:diffalpha}
  \bar{\alpha}_{k+1}^+(z) - \bar{\alpha}_k^+(z) =\\
  \frac{(s_{0,k-1}-s_{0,k}) h_X(z) + (s_{0,k-1}^2 - s_{0,k-2}s_{0,k} ) h_{-BU}(z)}{s_{0,k}s_{0,k-1} h_{\bar{W}}(z)}.
\end{multline}
\fi
  The denominator here is strictly positive, since $\lambda_i > 0$ and $s_{0,j}(\lambda_i)$ for all $j\in\mbb{N}_+$. In the numerator, 
  \begin{equation}
    s_{0,k-1} - s_{0,k} = -\lambda_i^k
  \end{equation}
  and
  \begin{equation}
    \begin{split}
      s_{0,k-2}s_{0,k} &=   (s_{0,k-1} - \lambda_i^{k-1})(s_{0,k-1} + \lambda_i^k)\\
      &=  s^2_{0,k-1} + (\lambda_i^k - \lambda_i^{k-1})s_{0,k-1} - \lambda_i^{k}\lambda^{k-1}_i
      \end{split}
    \end{equation}
    so
    \begin{equation}
      \begin{split}
        s_{0,k-1}^2 - s_{0,k-2}s_{0,k} &= \lambda_i^{k}\lambda^{k-1}_i - (\lambda_i^k - \lambda_i^{k-1})s_{0,k-1}\\
        &= \lambda_i^{k}\left( \lambda^{k-1}_i - (1 - \lambda^{-1}_i)s_{0,k-1} \right).
        \end{split}
      \end{equation}
      The term within parentheses may be written
      \begin{equation}
        \begin{split}
          \lambda^{k-1}_i - (1 - \lambda^{-1}_i) s_{0,k-1} &= \lambda^{k-1}_i\\&\quad - (\lambda_i^0 + \lambda_i^1 + \dots + \lambda_i^{k-1})\\
          &\quad + (\lambda_i^{-1} + \lambda_i^0 + \dots + \lambda_i^{k-2})\\
          &= \lambda_i^{-1}.
          \end{split}
        \end{equation}
      The numerator of~\eqref{eq:diffalpha} is therefore, for $z=+\varphi_{i,1}$ or $-\varphi_{i,1}$,
      \begin{equation}
        -\lambda_i^k h_X(z)  + \lambda_i^k \lambda_i^{-1} h_{-BU}(z),
        \end{equation}
        which is negative if $\lambda_i h_X(z) > h_{-BU}(z)$, positive if $\lambda_i h_X(z) < h_{-BU}(z)$, and zero if $\lambda_i h_X(z) = h_{-BU}(z)$. Therefore, both $\{ {\alpha}_k^+(+\varphi_{i,1})\}_{k\in\mbb{N}_+}$ and $\{ {\alpha}_k^+(-\varphi_{i,1})\}_{k\in\mbb{N}_+}$ are positive sequences that are either monotonically decreasing, monotonically increasing, or constant, according to respective conditions. Consequently, the greatest lower bound of the sequence $\{ {\alpha}_k^+(z)\}_{k\in\mbb{N}_+}$, for $z = +\varphi_{i,1}$ or $-\varphi_{i,1}$, is $\lim_{k \to \infty} \alpha^+_k(z)$ if the sequence is decreasing, and $\alpha^+_1(\varphi_{i,1})$ otherwise. 
        \end{IEEEproof}
  
  The main ramification of Proposition~\ref{prop:alphaseq} is the following result.

  \begin{thm}\label{thm:inf}
  Consider the $ith$ Jordan block $J_i$ and suppose that the
  associated eigenvalue $\lambda_i$ is real and positive. If
  \begin{align}
    \lambda_i h_X(+\varphi_{i,1}) &> h_{-BU}(+\varphi_{i,1}) \\ \text{and} \ \lambda_i h_X(-\varphi_{i,1})&> h_{-BU}(-\varphi_{i,1}),
    \end{align}
    then
  \begin{equation}
    \inf_{k \in \mbb{N}_+} \bar{\alpha}_k^+(\lambda_i) = \max \{ \alpha^+_\infty(+\varphi_{i,1}), \alpha^+_\infty(-\varphi_{i,1})\} 
  \end{equation}%
  where, for $z = \pm\varphi_{i,1}$,
  \begin{equation}\label{eq:infbound}%
    {\alpha}_{\infty}^+(z) =
    \begin{cases}
      \textstyle\left(\frac{1}{\lambda_i}\right) \frac{h_{-BU}(z)}{h_{\bar{W}}(z)}, &  \lambda_i \geq 1, \\
      \textstyle(1-\lambda_i)
      \frac{h_{X}(z)}{h_{\bar{W}}(z)} +
      \frac{h_{-BU}(z)}{h_{\bar{W}}(z)}, & \lambda_i < 1.
    \end{cases}
  \end{equation}
  Consequently, if $\alpha > \inf_{k \in \mbb{N}_+} \bar{\alpha}_k^+(\lambda_i)$ then $C_\infty^\alpha = \emptyset$.
\end{thm}

\begin{IEEEproof}
  Under the stated conditions, the sequences $\{\alpha^+_k(+\varphi_{i,1})\}$ and $\{\alpha^+_k(-\varphi_{i,1})\}$, defined
by~\eqref{eq:alpha1}, are, for any $\lambda_i > 0$, positive and
monotonically decreasing with $k \in \mbb{N}_+$. Thus,
$\lim_{k \to
  \infty} {\alpha}^+_k(+\varphi_{i,1}) $ and $\lim_{k \to
  \infty} {\alpha}^+_k(-\varphi_{i,1}) $ exist, and are given by
\ifCLASSOPTIONonecolumn
\begin{equation}
  \alpha_\infty^+(\pm\varphi_{i,1}) \coloneqq \lim_{k \to \infty} {\alpha}_{k}(\pm\varphi_{i,1}) = \begin{cases} (1-\lambda_i) \frac{h_{X}(\pm\varphi_{i,1})}{h_{\bar{W}}(\pm\varphi_{i,1})} + \frac{h_{-BU}(\pm\varphi_{i,1})}{h_{\bar{W}}(\pm\varphi_{i,1})}, & 0 \leq \lambda_i < 1, \\
    \frac{h_{-BU}(\pm\varphi_{i,1})}{h_{\bar{W}}(\pm\varphi_{i,1})}, & \lambda_i = 1, \\
    \left(\frac{1}{\lambda_i}\right)\frac{h_{-BU}(\pm\varphi_{i,1})}{h_{\bar{W}}(\pm\varphi_{i,1})}, & \lambda_i > 1.\end{cases}
\end{equation}
\else
\begin{multline}
  \alpha_\infty^+(\pm\varphi_{i,1}) \coloneqq \lim_{k \to \infty} {\alpha}_{k}(\pm\varphi_{i,1}) = \\\begin{cases} (1-\lambda_i) \frac{h_{X}(\pm\varphi_{i,1})}{h_{\bar{W}}(\pm\varphi_{i,1})} + \frac{h_{-BU}(\pm\varphi_{i,1})}{h_{\bar{W}}(\pm\varphi_{i,1})}, & 0 \leq \lambda_i < 1, \\
    \frac{h_{-BU}(\pm\varphi_{i,1})}{h_{\bar{W}}(\pm\varphi_{i,1})}, & \lambda_i = 1, \\
    \left(\frac{1}{\lambda_i}\right)\frac{h_{-BU}(\pm\varphi_{i,1})}{h_{\bar{W}}(\pm\varphi_{i,1})}, & \lambda_i > 1.\end{cases}
\end{multline}
\fi
  The $\lambda_i = 1$ case is subsumed by the $\abs{\lambda_i} > 1$ case, since the same expression applies evaluated at $\lambda_i = 1$.

  If $h_{S^\alpha_\infty}(+\varphi_{i,1}) < 0$ and $h_{S^\alpha_\infty}(-\varphi_{i,1}) < 0 $ then $S_\infty^\alpha = \emptyset \implies C_\infty^\alpha = \emptyset$. Thus,
  \begin{equation}
    \alpha > \max\{ \alpha_{\infty}(+\varphi_{i,1}), \alpha_\infty(-\varphi_{i,1})\}
    \end{equation}
    is sufficient to ensure $C_\infty^\alpha = \emptyset.$
    \end{IEEEproof}

The result provides clear and novel confirmation of the obvious and
well-known fact that non-existence of a robust control law and control
invariant set is more easily established for unstable $A$. In
particular, when $A$ has an unstable real positive eigenvalue, the
support function of the disturbance set $\alpha\bar{W}$ in the
direction $\pm \varphi_{i,1}$ just needs to exceed that of the mapped
input set $BU$ in the opposite direction $\mp \varphi_{i,1}$ reduced by
the magnitude of $\lambda_i > 1$. Interestingly, the $\alpha$
sufficient to cause emptiness of $C^{\alpha}_\infty$ is dependent on
the relative size of $\bar{W}$ to $X$ for stable $A$, but completely
independent of the relative sizes of these sets for unstable $A$.
  
      \begin{rem}
        An obvious strategy for the malicious attack of an unstable
        system is to select the disturbance $w \in W=\alpha \bar{W}$
        as $w(x,u) = -Bu(x)$ at every $x$, should the information
        pattern permit this. Clearly, this nullifies the effect of the
        control input and means the state leaves $X$ in finite
        time. This attack strategy is related to the so-called DoS
        attacks or jamming attacks, where the transmission of the
        control signal from the controller to the actuator is
        blocked. In~\cite{PT15}, a deterministic model of such attacks
        has been proposed, which takes account of constraints on the
        duration and frequency for launching such attacks; conditions
        for maintaining stability of networked control systems are
        obtained.

        For this attack strategy to be possible, $w(x,u) = -Bu(x)$ has to be admissible for all $x\in X$,
        which in turn requires that $\alpha h_{\bar{W}}(z) \geq h_{-BU}(z)$
        for all $z\in\mbb{R}^n$. Bound~\eqref{eq:infbound} reduces this value
        of critical $\alpha$ by a factor of $\lambda_i > 1$,
        \emph{disallowing} the strategy $w(x,u) = -Bu(x)$ but still
        guaranteeing non-existence of $C^\alpha_{\infty}$.

        \end{rem}

        \section{General eigenvalue cases}
\label{sec:general}

        Building upon the preceding analysis that developed conditions
        for the emptiness of $C_\infty^\alpha$ by considering real
        positive eigenvalues and their associated eigenvectors, this
        section introduces a novel and technically significant
        extension to our general framework, addressing previously
        unexamined complexities. A comprehensive understanding of
        robust control policy existence requires considering all
        possible eigenvalue cases, including repeated eigenvalues and
        particularly real-negative and complex eigenvalues. These
        distinct eigenvalue types introduce unique characteristics to
        the system's spectral properties and their influence on the
        critical disturbance size, and their detailed analysis within
        our established framework represents a key original
        contribution of this work. In this section, therefore, we
        derive conditions for the emptiness of $C^\alpha_\infty$ that
        account for these different eigenvalue types and behaviors,
        offering new and insightful perspectives.

        \ifextended

        Longer proofs and those of technical lemmas in this section are provided in appendices.

        \fi

      \subsection{The eigenvalue $\lambda_i$ is real and positive---first generalized eigenvector}
\label{sec:geneigs}

The results of Section~\ref{sec:eig} can be augmented by considering
additional directions associated with $i$th Jordan block of $A^\top$, $J_i$, when
its size $n_i > 1$. This applies to the situation of a system with
repeated, possibly dominant, eigenvalues, such as the double
integrator.

The main challenge with extending the previous analysis stems from
the coupling between generalized eigenvectors induced by the Jordan chain,
and the sublinearity property of the support function. Consider the
case where $n_i = 2$, \ie~the real, positive eigenvalue $\lambda_i$ has an algebraic multiplicity of at least two, and the Jordan block $J_i$ is $2\times 2$; the eigenvector is $\varphi_{i,1}$ and the generalized eigenvector is $\varphi_{i,2}$. The support function evaluation of~\eqref{eq:Scond2} in
the directions~$\pm \varphi_{i,2}$ yields
\ifCLASSOPTIONonecolumn
\begin{equation}\label{eq:34}
(\forall k \in \mbb{N})\ {h}_{S_k^{\alpha}}(\pm\varphi_{i,2}) \leq {h}_X(\pm\varphi_{i,2})  + \sum_{\ell=0}^{k-2} {h}_{A^\ell (-BU)}(\pm\varphi_{i,2}) - \sum_{\ell=0}^{k-1} {h}_{A^\ell (\alpha\bar{W})}(\pm\varphi_{i,2}).
\end{equation}
\else
  \begin{multline}\label{eq:34}
(\forall k \in \mbb{N})\ {h}_{S_k^{\alpha}}(\pm\varphi_{i,2}) \leq {h}_X(\pm\varphi_{i,2}) \\ + \sum_{\ell=0}^{k-2} {h}_{A^\ell (-BU)}(\pm\varphi_{i,2}) - \sum_{\ell=0}^{k-1} {h}_{A^\ell (\alpha\bar{W})}(\pm\varphi_{i,2}).
\end{multline}
\fi
The right-hand side of this inequality is, using~\eqref{eq:AD},
\ifCLASSOPTIONonecolumn
\begin{equation}\label{eq:35}
(\forall k \in \mbb{N})\ {h}_{X}(\pm \varphi_{i,2})  + \sum_{\ell=0}^{k-2} {h}_{-BU}\left( \pm(\lambda_i^\ell \varphi_{i,2} + \ell\lambda_i^{\ell-1}\varphi_{i,1}) \right) - \alpha \sum_{\ell=0}^{k-1} {h}_{\bar{W}}\left(\pm(\lambda_i^\ell \varphi_{i,2} + \ell\lambda_i^{\ell-1}\varphi_{i,1} ) \right).
\end{equation}
\else
\begin{multline}\label{eq:35}
(\forall k \in \mbb{N})\ {h}_{X}(\pm \varphi_{i,2})  + \sum_{\ell=0}^{k-2} {h}_{-BU}\left( \pm(\lambda_i^\ell \varphi_{i,2} + \ell\lambda_i^{\ell-1}\varphi_{i,1}) \right) \\- \alpha \sum_{\ell=0}^{k-1} {h}_{\bar{W}}\left(\pm(\lambda_i^\ell \varphi_{i,2} + \ell\lambda_i^{\ell-1}\varphi_{i,1} ) \right).
\end{multline}
\fi
Because the support function is sublinear, however, it is not
  possible, without further assumptions, to
  \emph{equate}~\eqref{eq:35} with expressions involving linear terms in
  $h_Y(\varphi_{i,2})$ and $h_Y(\varphi_{i,1})$. Moreover, it is not simple
  even to upper bound~\eqref{eq:35} with a view to obtaining a
  sufficient condition for emptiness of $S_k^\alpha$,
  because~\eqref{eq:35} involves a subtraction; a lower bound on the
  subtractive term is required. The following result helps provide
  such a lower bound.

  \begin{lem}\label{lem:hbound}
    Consider a set of points $z_i \in \mbb{R}^n, i\in\mc{M}\coloneq\{1,\dots,M\}$, and a C-set $Y \subset \mbb{R}^n$. The support function satisfies
\begin{equation}\label{eq:hbound}
    \max_{i\in\mc{M}} \{ h_Y(z_i) - \!\!\!\sum_{j\in\mc{M}\setminus\{i\}} \!\!\! h_{-Y}(z_j) \} 
    \leq h_Y\Big(\sum_{i\in\mc{M}} z_i\Big) \leq \sum_{i\in\mc{M}} h_Y(z_i).
\end{equation}

  \end{lem}
  We also make the following simplifying assumption; though the
  analysis that follows could be done without this assumption, the
  derived bounds would be far more complicated.

  \begin{assum}\label{assump:W}
    The set $\bar{W}$ is symmetrical about the origin.
  \end{assum}

  \ifCLASSOPTIONonecolumn
\begin{figure*}[p]
      \footnotesize
      \hrulefill
\vspace*{4pt}
    \setcounter{equation}{177}
\begin{equation}
      \beta^+_k(\pm\varphi_{i,2}) =  \begin{cases} 
        \frac{h_X(\pm\varphi_{i,2}) + \left( \frac{1-\lambda_i^{k-1}}{1-\lambda_i} \right)h_{-BU}(\pm\varphi_{i,2}) + \left(\frac{1-(k-1)\lambda_i^{k-2} + (k-2)\lambda_i^{k-1}}{(1-\lambda_i)^2}\right) h_{-BU}(\pm\varphi_{i,1})}{\left(\frac{1-\lambda_i^{k}}{1-\lambda_i}\right)  h_{\bar{W}}(\varphi_{i,2}) - \left( \frac{1- k \lambda_i^{k-1} + (k-1)\lambda_i^k}{(1-\lambda_i)^2} \right) h_{\bar{W}}(\varphi_{i,1})}, & 1 \leq k \leq \floor*{\ell_i^*}+1, \lambda_{i} \neq 1, \\[1em]
        \frac{h_X(\pm\varphi_{i,2}) + (k-1) h_{-BU}(\pm\varphi_{i,2}) + \left(\frac{(k-1)(k-2)}{2}\right) h_{-BU}(\pm\varphi_{i,1})}{k h_{\bar{W}}(\varphi_{i,2}) - \left( \frac{k(k-1)}{2} \right)  h_{\bar{W}}(\varphi_{i,1})}, & 1 \leq k \leq \floor*{\ell_i^*}+1, \lambda_{i} = 1, \\[1em]
     \frac{h_X(\pm\varphi_{i,2}) + \Bigl( \frac{1-\lambda_i^{k-1}}{1-\lambda_i} \Bigr)h_{-BU}(\pm\varphi_{i,2}) + \Bigl(\frac{1-(k-1)\lambda_i^{k-2} + (k-2)\lambda_i^{k-1}}{(1-\lambda_i)^2}\Bigr) h_{-BU}(\pm\varphi_{i,1})}{\Bigl(\frac{1-2\lambda_i^{\floor*{\ell_i^*}+1}  + \lambda_i^k}{1-\lambda_i}\Bigr) h_{\bar{W}}(\varphi_{i,2}) - \Bigl( \frac{1 -2(\floor*{\ell_i^*}+1)\lambda_i^{\floor*{\ell_i^*}} + 2\floor*{\ell_i^*}\lambda_i^{\floor*{\ell_i^*}+1} - (k-1)\lambda_i^{k} + k \lambda_i^{k-1} }{(1-\lambda_i)^2} \Bigr) h_{\bar{W}}(\varphi_{i,1})}, & k > \floor*{\ell_i^*}+1,\lambda_{i} \neq 1, \\[1em]
        \frac{h_X(\pm\varphi_{i,2}) + (k-1) h_{-BU}(\pm\varphi_{i,2}) + \left(\frac{(k-1)(k-2)}{2}\right) h_{-BU}(\pm\varphi_{i,1})}{ \left(2(\floor*{\ell_i^*}+1)-k \right) h_{\bar{W}}(\varphi_{i,2}) + \left( \frac{k^2- k - 2\floor*{\ell_i^*}^2 - 2\floor*{\ell_i^*}}{2}  \right)  h_{\bar{W}}(\varphi_{i,1})}, & k > \floor*{\ell_i^*}+1, \lambda_{i} = 1. \end{cases}\label{eq:alpha2}
    \end{equation}%
\begin{equation}
  {\alpha}^-_k(\pm\varphi_{i,1}) =  \begin{cases} 
    \frac{h_X(\pm\varphi_{i,1}) + \abs{\lambda_i}\biggl( \frac{1-\abs{\lambda_i}^{2\floor*{\frac{k-1}{2}}}}{1-\abs{\lambda_i}^2} \biggr)h_{-BU}(\mp\varphi_{i,1}) + \biggl( \frac{1-\abs{\lambda_i}^{2\ceil*{\frac{k-1}{2}}}}{1-\abs{\lambda_i}^2} \biggr)h_{-BU}(\pm\varphi_{i,1})}{\abs{\lambda_i}\biggl( \frac{1-\abs{\lambda_i}^{2\floor*{\frac{k}{2}}}}{1-\abs{\lambda_i}^2} \biggr)h_{\bar{W}}(\mp\varphi_{i,1}) + \biggl( \frac{1-\abs{\lambda_i}^{2\ceil*{\frac{k}{2}}}}{1-\abs{\lambda_i}^2} \biggr)h_{\bar{W}}(\pm\varphi_{i,1})}, & \lambda_{i} \neq -1, \\
    \frac{h_X(\pm\varphi_{i,1}) + \floor*{\frac{k-1}{2}} h_{-BU}(\mp\varphi_{i,1}) + \ceil*{\frac{k-1}{2}} h_{-BU}(\pm\varphi_{i,1})}{ \floor*{\frac{k}{2}} h_{\bar{W}}(\mp\varphi_{i,1}) +   \ceil*{\frac{k}{2}}h_{\bar{W}}(\pm\varphi_{i,1})}, & \lambda_{i} = -1. \end{cases} \label{eq:alphaminus} 
\end{equation}
\begin{align}
  \lim_{n \to \infty} {\alpha}^-_{2n+1} (\pm\varphi_{i,1}) &=  \begin{cases} \frac{( 1-\abs{\lambda_i}^2) h_X(\pm\varphi_{i,1}) + \abs{\lambda_i} h_{-BU}(\mp\varphi_{i,1}) + h_{-BU}(\pm\varphi_{i,1})}{ \abs{\lambda_i} h_{\bar{W}}(\mp\varphi_{i,1}) + h_{\bar{W}}(\pm\varphi_{i,1})}, & 0 < \abs{\lambda_{i}} < 1, \\
    \frac{\abs{\lambda_i} h_{-BU}(-\varphi_{i,1}) + h_{-BU}(\varphi_{i,1}) }{\abs{\lambda_i} h_{\bar{W}}(-\varphi_{i,1}) + \abs{\lambda_i}^2 h_{\bar{W}}(\varphi_{i,1})},  & \abs{\lambda_{i}} \geq 1. \end{cases} \label{eq:alphaminusinf}\\
  \lim_{n \to \infty} {\alpha}^-_{2n} (\pm\varphi_{i,1}) &=  \begin{cases} \frac{( 1-\abs{\lambda_i}^2) h_X(\pm\varphi_{i,1}) + \abs{\lambda_i} h_{-BU}(\mp\varphi_{i,1}) + h_{-BU}(\pm\varphi_{i,1})}{ \abs{\lambda_i} h_{\bar{W}}(\mp\varphi_{i,1}) + h_{\bar{W}}(\pm\varphi_{i,1})}, & 0 < \abs{\lambda_{i}} < 1, \\
   \frac{ h_{-BU}(\mp\varphi_{i,1}) + \abs{\lambda_i} h_{-BU}(\pm\varphi_{i,1}) }{\abs{\lambda_i}^2 h_{\bar{W}}(\mp\varphi_{i,1}) + \abs{\lambda_i} h_{\bar{W}}(\pm\varphi_{i,1})} ,  & \abs{\lambda_{i}} \geq 1. \end{cases} \label{eq:alphaminusinf2}
\end{align}

\setcounter{equation}{59}
\footnotesize
\end{figure*}
  \else
\begin{figure*}[!b]
      \normalsize
      \hrulefill
\vspace*{4pt}
    \newcounter{MYtempeqncnt2}
    \setcounter{MYtempeqncnt2}{\value{equation}}
    \setcounter{equation}{76}
\begin{equation}
      \beta^+_k(\pm\varphi_{i,2}) =  \begin{cases} 
        \frac{h_X(\pm\varphi_{i,2}) + \left( \frac{1-\lambda_i^{k-1}}{1-\lambda_i} \right)h_{-BU}(\pm\varphi_{i,2}) + \left(\frac{1-(k-1)\lambda_i^{k-2} + (k-2)\lambda_i^{k-1}}{(1-\lambda_i)^2}\right) h_{-BU}(\pm\varphi_{i,1})}{\left(\frac{1-\lambda_i^{k}}{1-\lambda_i}\right)  h_{\bar{W}}(\varphi_{i,2}) - \left( \frac{1- k \lambda_i^{k-1} + (k-1)\lambda_i^k}{(1-\lambda_i)^2} \right) h_{\bar{W}}(\varphi_{i,1})}, & 1 \leq k \leq \floor*{\ell_i^*}+1, \lambda_{i} \neq 1, \\[1em]
        \frac{h_X(\pm\varphi_{i,2}) + (k-1) h_{-BU}(\pm\varphi_{i,2}) + \left(\frac{(k-1)(k-2)}{2}\right) h_{-BU}(\pm\varphi_{i,1})}{k h_{\bar{W}}(\varphi_{i,2}) - \left( \frac{k(k-1)}{2} \right)  h_{\bar{W}}(\varphi_{i,1})}, & 1 \leq k \leq \floor*{\ell_i^*}+1, \lambda_{i} = 1, \\[1em]
     \frac{h_X(\pm\varphi_{i,2}) + \Bigl( \frac{1-\lambda_i^{k-1}}{1-\lambda_i} \Bigr)h_{-BU}(\pm\varphi_{i,2}) + \Bigl(\frac{1-(k-1)\lambda_i^{k-2} + (k-2)\lambda_i^{k-1}}{(1-\lambda_i)^2}\Bigr) h_{-BU}(\pm\varphi_{i,1})}{\Bigl(\frac{1-2\lambda_i^{\floor*{\ell_i^*}+1}  + \lambda_i^k}{1-\lambda_i}\Bigr) h_{\bar{W}}(\varphi_{i,2}) - \Bigl( \frac{1 -2(\floor*{\ell_i^*}+1)\lambda_i^{\floor*{\ell_i^*}} + 2\floor*{\ell_i^*}\lambda_i^{\floor*{\ell_i^*}+1} - (k-1)\lambda_i^{k} + k \lambda_i^{k-1} }{(1-\lambda_i)^2} \Bigr) h_{\bar{W}}(\varphi_{i,1})}, & k > \floor*{\ell_i^*}+1,\lambda_{i} \neq 1, \\[1em]
        \frac{h_X(\pm\varphi_{i,2}) + (k-1) h_{-BU}(\pm\varphi_{i,2}) + \left(\frac{(k-1)(k-2)}{2}\right) h_{-BU}(\pm\varphi_{i,1})}{ \left(2(\floor*{\ell_i^*}+1)-k \right) h_{\bar{W}}(\varphi_{i,2}) + \left( \frac{k^2- k - 2\floor*{\ell_i^*}^2 - 2\floor*{\ell_i^*}}{2}  \right)  h_{\bar{W}}(\varphi_{i,1})}, & k > \floor*{\ell_i^*}+1, \lambda_{i} = 1. \end{cases}\label{eq:alpha2}
    \end{equation}%
\begin{equation}
  {\alpha}^-_k(\pm\varphi_{i,1}) =  \begin{cases} 
    \frac{h_X(\pm\varphi_{i,1}) + \abs{\lambda_i}\biggl( \frac{1-\abs{\lambda_i}^{2\floor*{\frac{k-1}{2}}}}{1-\abs{\lambda_i}^2} \biggr)h_{-BU}(\mp\varphi_{i,1}) + \biggl( \frac{1-\abs{\lambda_i}^{2\ceil*{\frac{k-1}{2}}}}{1-\abs{\lambda_i}^2} \biggr)h_{-BU}(\pm\varphi_{i,1})}{\abs{\lambda_i}\biggl( \frac{1-\abs{\lambda_i}^{2\floor*{\frac{k}{2}}}}{1-\abs{\lambda_i}^2} \biggr)h_{\bar{W}}(\mp\varphi_{i,1}) + \biggl( \frac{1-\abs{\lambda_i}^{2\ceil*{\frac{k}{2}}}}{1-\abs{\lambda_i}^2} \biggr)h_{\bar{W}}(\pm\varphi_{i,1})}, & \lambda_{i} \neq -1, \\
    \frac{h_X(\pm\varphi_{i,1}) + \floor*{\frac{k-1}{2}} h_{-BU}(\mp\varphi_{i,1}) + \ceil*{\frac{k-1}{2}} h_{-BU}(\pm\varphi_{i,1})}{ \floor*{\frac{k}{2}} h_{\bar{W}}(\mp\varphi_{i,1}) +   \ceil*{\frac{k}{2}}h_{\bar{W}}(\pm\varphi_{i,1})}, & \lambda_{i} = -1. \end{cases} \label{eq:alphaminus} 
\end{equation}
\begin{align}
  \lim_{n \to \infty} {\alpha}^-_{2n+1} (\pm\varphi_{i,1}) &=  \begin{cases} \frac{( 1-\abs{\lambda_i}^2) h_X(\pm\varphi_{i,1}) + \abs{\lambda_i} h_{-BU}(\mp\varphi_{i,1}) + h_{-BU}(\pm\varphi_{i,1})}{ \abs{\lambda_i} h_{\bar{W}}(\mp\varphi_{i,1}) + h_{\bar{W}}(\pm\varphi_{i,1})}, & 0 < \abs{\lambda_{i}} < 1, \\
    \frac{\abs{\lambda_i} h_{-BU}(-\varphi_{i,1}) + h_{-BU}(\varphi_{i,1}) }{\abs{\lambda_i} h_{\bar{W}}(-\varphi_{i,1}) + \abs{\lambda_i}^2 h_{\bar{W}}(\varphi_{i,1})},  & \abs{\lambda_{i}} \geq 1. \end{cases} \label{eq:alphaminusinf}\\
  \lim_{n \to \infty} {\alpha}^-_{2n} (\pm\varphi_{i,1}) &=  \begin{cases} \frac{( 1-\abs{\lambda_i}^2) h_X(\pm\varphi_{i,1}) + \abs{\lambda_i} h_{-BU}(\mp\varphi_{i,1}) + h_{-BU}(\pm\varphi_{i,1})}{ \abs{\lambda_i} h_{\bar{W}}(\mp\varphi_{i,1}) + h_{\bar{W}}(\pm\varphi_{i,1})}, & 0 < \abs{\lambda_{i}} < 1, \\
   \frac{ h_{-BU}(\mp\varphi_{i,1}) + \abs{\lambda_i} h_{-BU}(\pm\varphi_{i,1}) }{\abs{\lambda_i}^2 h_{\bar{W}}(\mp\varphi_{i,1}) + \abs{\lambda_i} h_{\bar{W}}(\pm\varphi_{i,1})} ,  & \abs{\lambda_{i}} \geq 1. \end{cases} \label{eq:alphaminusinf2}
\end{align}

\setcounter{equation}{\value{MYtempeqncnt2}}
\footnotesize
\end{figure*}
\fi

  Applying Lemma~\ref{lem:hbound} to the terms in~\eqref{eq:35} yields, for the additive term,
  \ifCLASSOPTIONonecolumn
  \begin{equation}\label{eq:tbound}
    (\forall k \in \mbb{N}) \ \sum_{\ell=0}^{k-2} {h}_{-BU}\left(\pm(\lambda_i^\ell\varphi_{i,2} + \ell\lambda_i^{\ell-1}\varphi_{i,1}) \right)  \leq  \sum_{\ell=0}^{k-2} \lambda_{i}^\ell  h_{- BU}(\pm\varphi_{i,2}) + \sum_{\ell=0}^{k-2} \ell \lambda_{i}^{\ell-1}  h_{- BU}(\pm\varphi_{i,1}),
  \end{equation}
  \else
  \begin{multline}\label{eq:tbound}
    (\forall k \in \mbb{N}) \ \sum_{\ell=0}^{k-2} {h}_{-BU}\left(\pm(\lambda_i^\ell\varphi_{i,2} + \ell\lambda_i^{\ell-1}\varphi_{i,1}) \right) \\ \leq  \sum_{\ell=0}^{k-2} \lambda_{i}^\ell  h_{- BU}(\pm\varphi_{i,2}) + \sum_{\ell=0}^{k-2} \ell \lambda_{i}^{\ell-1}  h_{- BU}(\pm\varphi_{i,1}),
  \end{multline}
  \fi
    and, for the subtractive term, in view of Assumption~\ref{assump:W},
    \begin{equation}\label{eq:bbound}
      \sum_{\ell=0}^{k-1} {h}_{\bar{W}}\left( \lambda_i^\ell\varphi_{i,2} + \ell\lambda_i^{\ell-1}\varphi_{i,1} \right) \geq  \sum_{\ell=0}^{k-1} \max\{ f_i(\ell), -f_i(\ell) \},
      \end{equation}
      where,
      \begin{equation}\label{eq:f}
       (\forall \ell\in\mbb{N})\ f_i(\ell)\coloneqq
    \lambda_i^\ell h_{\bar{W}}(\varphi_{i,2}) - \ell\lambda_i^{\ell-1}h_{\bar{W}}(\varphi_{i,1}).
    \end{equation}
    The following result establishes some properties of the function $f_i$ that are instrumental in deriving a counterpart to Theorem~\ref{thm:main} corresponding to the direction $z=\varphi_{i,2}$. Let
    \begin{equation}\label{eq:ell}
      \ell^*_i \coloneqq \lambda_i \frac{h_{\bar{W}}(\varphi_{i,2})}{h_{\bar{W}}(\varphi_{i,1})} > 0.
      \end{equation}

    \begin{lem}\label{lem:f}
      The function $f_i(\ell)$ satisfies the following properties: (i) $f_i(0) = h_{\bar{W}}(\varphi_{i,2})$; (ii) $f_i(\ell) = 0$ if and only if $\ell = \ell_i^*$; (iii) if $0 \leq \ell < \ell_i^*$ then $f_i(\ell) > 0$; (iv) if $\ell > \ell_i^*$ then $f_i(\ell) < 0$; (v)
      \begin{equation}
        \max \{ f_i(\ell),-f_i(\ell)\}= \begin{dcases} +f_i(\ell) > 0, & 0 \leq \ell < \ell^*_i,\\
          -f_i(\ell) > 0, & \ell > \ell^*_{i}.\end{dcases}
      \end{equation}
      Finally, (vi),
      \begin{equation}
        (\forall k \in \mbb{N}_+) \ \sum_{\ell=0}^{k-1} \max\{ f_i(\ell), -f_i(\ell) \} > 0 .
        \end{equation}
    
    \end{lem}

    Lemma~\ref{lem:f} ensures that the lower bound
    in~\eqref{eq:bbound} is well defined and positive. The next result
    then uses the bounds~\eqref{eq:tbound} and~\eqref{eq:bbound} to
    determine the $\alpha$ sufficient to ensure emptiness of $C_k^\alpha$. A proof of the result is provided in \ifextended Appendix~\ref{proof:3} \fi.

    \begin{thm}\label{thm:3}
      Consider the $ith$ Jordan block $J_i$, and suppose that its size
      $n_i \geq 2$, the associated eigenvalue $\lambda_i$ is real and
      positive, and Assumption~\ref{assump:W} holds. If, for some
      $k^\star\in\mbb{N}_+$,
    \begin{equation}
      \alpha > \bar{\beta}^+_{k^\star}(\lambda_i) \coloneqq \max\{ {\beta}^{+}_{k^\star}(+\varphi_{i,2}), \beta^{+}_{k^\star}(-\varphi_{i,2}) \},
    \end{equation}
    where $\beta^+_{k}(\pm \varphi_{i,2})$, for $k\in\mbb{N}_+$, is defined in~\eqref{eq:alpha2}, then $C_k^\alpha = \emptyset$ for all $k\geq k^\star$. 
    \end{thm}

    \begin{rem}
      To develop a counterpart to Theorem~\ref{thm:inf} 
      $\inf_{k \in \mbb{N}} \bar{\beta}^+_{k^\star}(\lambda_i) $ needs to be evaluated. However,
      the sequences $\{\beta_k^+(+ \varphi_{i,2})\}$ and $\{ \beta_k^+(-\varphi_{i,1})\}$ are, in general,
      non-monotonic with $k\in\mbb{N}_+$ (see Fig.~\ref{fig:double}), so
      $\inf_{k \in \mbb{N}} \beta_k^+(\pm \varphi_{i,2})$ is, in general, neither equal to $\lim_{k \to \infty} \beta_k^+(\pm \varphi_{i,2})$ nor equal to $\beta_1^+(\pm\varphi_{i,2})$. Moreover,
      analysing the minima of $\{\beta_k^+(\pm\varphi_{i,2})\}$ is a challenging task with a complex dependence on the system eigenvalues and constraint sets.
      \end{rem}

\begin{rem}      
  Further, but more involved, sufficient conditions could be obtained
  for the case where the size of the Jordan block exceeds two by
  considering the directions $\varphi_{i,j}, j = 3, 4, \dots, n_i$,
  \ie~those of the second and higher generalized eigenvectors
  associated with the repeated real positive eigenvalue
  $\lambda_i$. The sublinearity of the support function,
  which prompted the additional analysis of Lemma~\ref{lem:hbound} and
  Lemma~\ref{lem:f}, poses a significant challenge, however, for it will be required to analyse the behaviour of a lower bound on
  \begin{equation}
    h_Y\left( \sum_{p=0}^{j-1} \binom{\ell}{p} \varphi_{i,j-p} \right).
    \end{equation}
\end{rem}
  
  \subsection{The eigenvalue $\lambda_i$ is real and negative}

   If the real eigenvalue $\lambda_i$ associated with Jordan block $J_i$ is negative, then evaluation of the support function $h_{A^\ell Y}(z)$ for $z=\pm\varphi_{i,1}$ and $\ell = 0,1,\dots$ yields
  \begin{equation}\label{eq:hneg}
h_{A^\ell Y}(\pm\varphi_{i,1}) =  h_{Y}(\pm \lambda_i^\ell \varphi_{i,1}) = \begin{dcases} \abs{\lambda_i}^\ell h_{Y}( \pm\varphi_{i,1}), & \ell \ \text{even}, \\ \abs{\lambda_i}^\ell h_{Y}( \mp\varphi_{i,1}), & \ell \ \text{odd}. \end{dcases}
    \end{equation}
    That is, the support of a set $Y$ needs to be evaluated in opposite directions for odd and even values of $\ell$. This presents no conceptual barrier, but leads to more complicated expressions that split summations into odd and even terms.

    
\begin{thm}\label{thm:neg}
  Consider the $ith$ Jordan block $J_i$ and suppose that the
  associated eigenvalue $\lambda_i$ is real and negative. If, for some $k^\star\in\mbb{N}$,
    \begin{equation}
      \alpha > \bar{\alpha}_{k^\star}^-(\lambda_i) \coloneqq \max\{ \alpha^-_{k^\star}(+\varphi_{i,1}), \alpha^-_{k^\star}(-\varphi_{i,1}) \},
    \end{equation}
    where $\alpha^-_k(\pm \varphi_{i,1})$, for $k\in\mbb{N}_+$, is defined in~\eqref{eq:alphaminus}, then $C^\alpha_k = \emptyset$ for all $k\geq k^\star$.
    \end{thm}

    \begin{IEEEproof}
      The proof proceeds in a similar way to that of Theorem~\ref{thm:main}, starting from~\eqref{eq:31} but noting~\eqref{eq:hneg} applies when $\lambda_i < 0$. Again, the development is given for the positive direction $z=+\varphi_{i,1}$ but is easily extended to the opposite direction via a change of sign in support function arguments.

     The support of $S_k^\alpha$ satisfies
     \ifCLASSOPTIONonecolumn
\begin{multline}
(\forall k \in \mbb{N})\ {h}_{S_k^{\alpha}}(\varphi_{i,1}) \leq {h}_X(\varphi_{i,1})  + {h}_{-BU}( -\varphi_{i,1}) \sum_{\substack{\ell=1\\\ell \, \text{odd}}}^{k-2} \abs{\lambda_i}^\ell + {h}_{-BU}( \varphi_{i,1}) \sum_{\substack{\ell=0\\\ell \, \text{even}}}^{k-2} \abs{\lambda_i}^\ell  \\ - \alpha {h}_{\bar{W}}(-\varphi_{i,1}) \sum_{\substack{\ell=1 \\ \ell \, \text{odd}}}^{k-1}\abs{\lambda_i}^\ell - \alpha {h}_{\bar{W}}(\varphi_{i,1}) \sum_{\substack{\ell=0 \\ \ell \, \text{even}}}^{k-1}\abs{\lambda_i}^\ell.
\end{multline}
     \else
     \begin{multline}
(\forall k \in \mbb{N})\ {h}_{S_k^{\alpha}}(\varphi_{i,1}) \leq {h}_X(\varphi_{i,1}) \\ + {h}_{-BU}( -\varphi_{i,1}) \sum_{\substack{\ell=1\\\ell \, \text{odd}}}^{k-2} \abs{\lambda_i}^\ell + {h}_{-BU}( \varphi_{i,1}) \sum_{\substack{\ell=0\\\ell \, \text{even}}}^{k-2} \abs{\lambda_i}^\ell  \\ - \alpha {h}_{\bar{W}}(-\varphi_{i,1}) \sum_{\substack{\ell=1 \\ \ell \, \text{odd}}}^{k-1}\abs{\lambda_i}^\ell - \alpha {h}_{\bar{W}}(\varphi_{i,1}) \sum_{\substack{\ell=0 \\ \ell \, \text{even}}}^{k-1}\abs{\lambda_i}^\ell.
\end{multline}
\fi
The result follows by using
\begin{align}
\sum_{\substack{\ell=1\\\ell \, \text{odd}}}^{k} \abs{\lambda_i}^\ell &= \sum_{\ell=0}^{\floor*{\frac{k-1}{2}}} \abs{\lambda_i}^{2\ell+1} = \begin{cases} \abs{\lambda_i}\frac{1-\abs{\lambda_i}^{2\floor*{\frac{k+1}{2}}}}{1-\abs{\lambda_i}^2}, & \abs{\lambda_i} \neq 1, \\ \floor*{\frac{k+1}{2}}, & \abs{\lambda_i} = 1,\end{cases} \\
\sum_{\substack{\ell=0\\\ell \, \text{even}}}^{k} \abs{\lambda_i}^\ell &= \sum_{\ell=0}^{\floor*{\frac{k}{2}}} \abs{\lambda_i}^{2\ell} = \begin{cases} \frac{1-\abs{\lambda_i}^{2\ceil*{\frac{k+1}{2}}}}{1-\abs{\lambda_i}^2}, & \abs{\lambda_i} \neq 1, \\ \ceil*{\frac{k+1}{2}}, & \abs{\lambda_i} = 1.\end{cases}
\end{align}
\end{IEEEproof}
    
It is then possible to characterize the smallest $\alpha$ sufficient to render $C_\infty^\alpha$ empty by, like in Section~\ref{sec:eig}, considering the behaviour of the sequences $\{\alpha^-_{k}(+\varphi_{i,1})\}_{k \in \mbb{N}_+}$ and $\{\alpha^-_{k}(-\varphi_{i,1})\}_{k \in \mbb{N}_+}$; in this case, there are two subsequences to be analysed, corresponding to odd and even values of $k$. A proof of the following technical result may be found in \ifextended Appendix~\ref{proof:alphaseqneg} \fi.

\begin{prop}\label{prop:alphaseqneg}
  The sequence~$\{{\alpha}^-_{k}(z)\}_{k \in \mbb{N}_+}$, defined by~\eqref{eq:alphaminus} with $z=+\varphi_{i,1}$ or $z = -\varphi_{i,1}$, is positive and satisfies the following properties:
    \begin{enumerate}
  \item $\{{\alpha}^-_{2n+1}(z) \}_{n \in \mbb{N}}$ is monotonically decreasing if
\begin{equation}\label{eq:negcond1}
 h_X(z) >  \frac{\left( \abs{\lambda_i} h_{-BU}(-z) +  h_{-BU}(z)\right) h_{\bar{W}}(z)}{\abs{\lambda_i} h_{\bar{W}}(-z) +  \abs{\lambda_i}^2 h_{\bar{W}}(z)} ,
\end{equation}
and non-decreasing otherwise;
\item $\{{\alpha}^-_{2n}(z) \}_{n \in \mbb{N}_+}$ is monotonically decreasing if
\begin{equation}\label{eq:negcond2}
  h_X(z) >  \frac{\left( \abs{\lambda_i} h_{\bar{W}}(-z) +  h_{\bar{W}}(z)\right) h_{-BU}(-z)}{\abs{\lambda_i}^2 h_{\bar{W}}(-z) +  \abs{\lambda_i} h_{\bar{W}}(z)} ,
\end{equation}
and non-decreasing otherwise.
\end{enumerate}

  \end{prop}

The next result characterizes the smallest $\alpha$ sufficient to guarantee emptiness of $C_\infty^\alpha$ by considering the direction associated with a real negative eigenvalue. Appendix~\ref{proof:neginf} contains the proof.
  
  \begin{thm}\label{thm:neginf}
Consider the $ith$ Jordan block $J_i$ and suppose that the
associated eigenvalue $\lambda_i$ is real and negative. If~\eqref{eq:negcond1} and~\eqref{eq:negcond2} hold, for both $z=+\varphi_{i,1}$ and $z=-\varphi_{i,1}$, then
\begin{equation}
\inf_{k \in \mbb{N}_+} \bar{\alpha}_k^-(\lambda_i) =  \min \{ \bar{\alpha}^{-,\ts{odd}}_\infty(\lambda_i), \bar{\alpha}^{-,\ts{even}}_\infty(\lambda_{i}) \},
  \end{equation}
  where,
  \ifCLASSOPTIONonecolumn
\begin{equation}
  \bar{\alpha}^{-,\ts{odd}}_\infty(\lambda_i) \coloneqq  \max \left\{ \lim_{n\to\infty} {\alpha}^-_{2n+1}(+\varphi_{i,1}), \lim_{n\to\infty} {\alpha}^-_{2n+1}(-\varphi_{i,1}) \right\},
\end{equation}
and
\begin{equation}
  \bar{\alpha}^{-,\ts{even}}_\infty(\lambda_i) \coloneqq   \max \left\{ \lim_{n\to\infty} {\alpha}^-_{2n}(+\varphi_{i,1}), \lim_{n\to\infty} {\alpha}^-_{2n}(-\varphi_{i,1}) \right\},
\end{equation}
  \else
\begin{multline}
  \bar{\alpha}^{-,\ts{odd}}_\infty(\lambda_i) \coloneqq \\  \max \left\{ \lim_{n\to\infty} {\alpha}^-_{2n+1}(+\varphi_{i,1}), \lim_{n\to\infty} {\alpha}^-_{2n+1}(-\varphi_{i,1}) \right\},
\end{multline}
\addtocounter{equation}{4} and
\begin{multline}
  \bar{\alpha}^{-,\ts{even}}_\infty(\lambda_i) \coloneqq \\  \max \left\{ \lim_{n\to\infty} {\alpha}^-_{2n}(+\varphi_{i,1}), \lim_{n\to\infty} {\alpha}^-_{2n}(-\varphi_{i,1}) \right\},
\end{multline}
\fi
  with the limits explicitly obtained in~\eqref{eq:alphaminusinf} and~\eqref{eq:alphaminusinf2}. Consequently, if $\alpha > \inf_{k \in \mbb{N}_+} \bar{\alpha}_k^-(\lambda_i)$ then $C_\infty^\alpha = \emptyset$.

\end{thm}

  This result stands apart from the positive--real one because of the distinct
  behaviour that a negative--real eigenvalue induces, and which the
  result captures. Indeed, note that Theorem~\ref{thm:main} and
  Theorem~\ref{thm:inf} arise from consideration of the evolution of
  the sets $A^\ell(-BU)$ and $A^\ell\bar{W}$ in the direction
  $+\varphi_{i,1}$ (or $-\varphi_{i,1}$), which, in the long run, is
  growth or diminishment in the same direction. On the other hand,
  Theorem~\ref{thm:neg} and Theorem~\ref{thm:neginf} arise from the
  consideration that the system state evolves in alternating
  directions. This requires analyzing the evolution of the sets
  $A^\ell (-BU)$ and $A^\ell \bar{W}$ as they reflect back and forth across
  the origin, rather than evolving in a single direction.

  \begin{rem}
    Note that the expressions in Theorem~\ref{thm:neg} and
    Theorem~\ref{thm:neginf} reduce to those in Theorem~\ref{thm:main}
    and Theorem~\ref{thm:inf} (respectively) if the sets $X$, $U$ and
    $\bar{W}$ are symmetrical.
   \end{rem}

          \subsection{The eigenvalue $\lambda_i$ is complex}

          The final case to consider is where the Jordan block $J_i$
          is associated with a complex eigenvalue. Let
          \begin{equation}
      \Phi_i =
      \begin{bmatrix}
        \varphi_{i,1}&
        \varphi_{i,2}&
        \dots&
        \varphi_{i,2n_i-1} &
        \varphi_{i,2n_i}
        \end{bmatrix}
      \end{equation}
      be the matrix containing the $2n_i$ columns of the modal matrix
      $\Phi$ corresponding to the block $J_i$ of size $n_i$. If $\lambda_i$ is complex, then the first two columns
      in $\Phi_i$ have the property
      \begin{equation}\label{eq:suprop1}
          (A^\ell)^\top \begin{bmatrix} \varphi_{i,1} & \varphi_{i,2} \end{bmatrix} =\rho^\ell_i \begin{bmatrix} \varphi_{i,1} & \varphi_{i,2} \end{bmatrix} R(\ell\theta_i),
        \end{equation}
        where $R$ is the rotation matrix defined in~\eqref{eq:R}. Therefore,
        \begin{equation}\label{eq:suprop2}
          (A^\ell)^\top \begin{bmatrix} \varphi_{i,1} & \varphi_{i,2} \end{bmatrix} R^{-1}(\ell\theta_i)=\rho^\ell_i \begin{bmatrix} \varphi_{i,1} & \varphi_{i,2} \end{bmatrix},
          \end{equation}
\ie~
        \begin{align}\label{eq:Acomplex}
          (A^\ell)^\top (\cos(\ell\theta_i)\varphi_{i,1} - \sin(\ell\theta_i)\varphi_{i,2}) &= \rho_i^\ell\varphi_{i,1},\\
          (A^\ell)^\top (\cos(\ell\theta_i)\varphi_{i,2} + \sin(\ell\theta_i)\varphi_{i,1}) &= \rho_i^\ell\varphi_{i,2}.
        \end{align}
        In light of these relations, the main idea is not to assess the condition~\eqref{eq:supc} in a \emph{constant} direction $\varphi_{i,1}$ or $\varphi_{i,2}$ (and their opposites), but rather, starting from directions
        \begin{align}
         \psi_{i,1}^{\ell_0} &\coloneqq \cos(\ell_0\theta_i)\varphi_{i,1} - \sin(\ell_0\theta_i)\varphi_{i,2},\\
          \psi_{i,2}^{\ell_0} &\coloneqq \cos(\ell_0\theta_i)\varphi_{i,2} + \sin(\ell_0\theta_i)\varphi_{i,1},
          \end{align}
          where $\ell_0 \in \mbb{N}$, to assess it in the rotating directions
        \begin{align}
         \psi_{i,1}^{\ell_0+\ell} &= \cos((\ell_0+\ell)\theta_i)\varphi_{i,1} - \sin((\ell_0+\ell)\theta_i)\varphi_{i,2},\\
          \psi_{i,2}^{\ell_0+\ell} &= \cos((\ell_0+\ell)\theta_i)\varphi_{i,2} + \sin((\ell_0+\ell)\theta_i)\varphi_{i,1},
          \end{align}
          for $\ell \in \mbb{N}_+$. Because of~\eqref{eq:suprop1} and~\eqref{eq:suprop2}, the support function has the property, for $j \in \{1,2\}$,
        %
        \begin{equation}\label{eq:psisupport}
          h_{A^\ell Y}(\psi^{\ell_0+\ell}_{i,j})=h_Y((A^\ell)^\top \psi^{\ell_0+\ell}_{i,j}) = \rho_i^\ell h_Y(\psi^{\ell_0}_{i,j}),
\end{equation}
which yields, for the support function of $S^\alpha_k$,
\ifCLASSOPTIONonecolumn
\begin{equation}\label{eq:hpsi}
(\forall k \in \mbb{N}_+)\ {h}_{S_k^{\alpha}}(\psi^{\ell_0+k}_{i,j}) \leq {h}_X(\psi^{\ell_0+k}_{i,j})  + {h}_{-BU}( \psi^{\ell_0}_{i,j}) \sum_{\ell=0}^{k-2} \rho_i^\ell   - \alpha {h}_{\bar{W}}(\psi^{\ell_0}_{i,j}) \sum_{\ell=0}^{k-1}\rho_i^\ell .
\end{equation}
\else
\begin{multline}\label{eq:hpsi}
(\forall k \in \mbb{N}_+)\ {h}_{S_k^{\alpha}}(\psi^{\ell_0+k}_{i,j}) \leq {h}_X(\psi^{\ell_0+k}_{i,j}) \\ + {h}_{-BU}( \psi^{\ell_0}_{i,j}) \sum_{\ell=0}^{k-2} \rho_i^\ell   - \alpha {h}_{\bar{W}}(\psi^{\ell_0}_{i,j}) \sum_{\ell=0}^{k-1}\rho_i^\ell .
\end{multline}
\fi
The same result up to a sign difference arises from considering the opposite directions $\psi_{i,1}^{\ell_0+k}$ and $\psi_{i,2}^{\ell_0+k}$.

We are nearly ready to provide counterpart to Theorem~\ref{thm:main}, but the dependency of~\eqref{eq:hpsi} on the initial directions $\psi^{\ell_0}_{i,1}$ and $\psi^{\ell_0}_{i,2}$ is problematic. To avoid considering all possible $\ell_0 \in \mbb{N}$ we make the following assumption.
\begin{assum}\label{assump:periodic}
  The argument of the complex eigenvalue $\lambda_i$, $\theta_i$, is such that $\theta_i = \frac{a}{b} \pi$ for integers  $a, b > 0$.
\end{assum}

\begin{lem}
If Assumption~\ref{assump:periodic} holds, then $\psi^{\ell}_{i,1}$ and $\psi^{\ell}_{i,2}$ are periodic, with period
\begin{equation}
  M_i = \frac{2b}{\gcd(a,2b)}.
\end{equation}
  \end{lem}

  Now we are ready to state Theorem~\ref{thm:comp1}, wherein we find, for each $k \in \mbb{N}_+$, a critical value of $\alpha$ that makes $C_k^\alpha = \emptyset$ by optimizing over the eigenvector $j\in\{1,2\}$ and initial direction $\psi^\ell_{i,j}$, $\ell_0 \in \{0,\dots,M_i\}$, to find the smallest $\alpha$ that makes $h_{S_k}(+\psi^{\ell_0+k}_{i,j})$ and $h_{S_k}(-\psi^{\ell_0+k}_{i,j})$ negative.

\begin{thm}\label{thm:comp1}
  Consider the $ith$ Jordan block $J_i$ and suppose that the
  associated eigenvalue $\lambda_i$ is complex, with $\theta_i$ satisfying Assumption~\ref{assump:periodic}. If, for some $k^\star\in\mbb{N}_+$,
  \ifCLASSOPTIONonecolumn
\begin{equation}
    \alpha > \bar{\alpha}^c_{k^\star}(\lambda_i) \coloneqq  \min_{j \in \{1,2\}} \min_{\ell_0 \in \{0,\ldots,M_i-1\}} \max\{ \alpha^{c}_{k^\star}(+\psi_{i,j}^{\ell_0}), \alpha^{c}_{k^\star}(-\psi_{i,j}^{\ell_0})\},
  \end{equation}
  \else
  \begin{multline}
    \alpha > \bar{\alpha}^c_{k^\star}(\lambda_i) \coloneqq \\ \min_{j \in \{1,2\}} \min_{\ell_0 \in \{0,\ldots,M_i-1\}} \max\{ \alpha^{c}_{k^\star}(+\psi_{i,j}^{\ell_0}), \alpha^{c}_{k^\star}(-\psi_{i,j}^{\ell_0})\},
  \end{multline}
  \fi
  where, for $k\in \mbb{N}_+$, $j \in \{1,2\}$, $\ell_0 \in \{0,\dots,M_i-1\}$,
  \ifCLASSOPTIONonecolumn
\begin{equation}\label{eq:alphac}
    {\alpha}^c_{k}(\pm\psi^{\ell_0}_{i,j}) \coloneqq
    \begin{cases}
      \left(\frac{1-\rho_i}{1-\rho_i^{k}}\right)\frac{h_{X}(\pm\psi^{\ell_0+k}_{i,j})}{h_{\bar{W}}(\pm\psi^{\ell_0}_{i,j})} + \left(\frac{1-\rho_i^{k-1}}{1-\rho_i^{k}}\right)\frac{h_{-BU}(\pm\psi^{\ell_0}_{i,j})}{h_{\bar{W}}(\pm\psi^{\ell_0}_{i,j})}, & \rho_i \neq 1, \\
      \left(\frac{1}{k}\right)
      \frac{h_{X}(\pm\psi^{\ell_0+k}_{i,j})}{h_{\bar{W}}(\pm\psi^{\ell_0}_{i,j})} +
      \left(\frac{k-1}{k}\right)
      \frac{h_{-BU}(\pm\psi^{\ell_0}_{i,j})}{h_{\bar{W}}(\pm\psi^{\ell_0}_{i,j})},&
      \rho_i = 1,
    \end{cases}
  \end{equation}
  \else
    \begin{multline}\label{eq:alphac}
    {\alpha}^c_{k}(\pm\psi^{\ell_0}_{i,j}) \coloneqq\\
    \begin{cases}
       \left(\frac{1-\rho_i}{1-\rho_i^{k}}\right)\frac{h_{X}(\pm\psi^{\ell_0+k}_{i,j})}{h_{\bar{W}}(\pm\psi^{\ell_0}_{i,j})} + \left(\frac{1-\rho_i^{k-1}}{1-\rho_i^{k}}\right)\frac{h_{-BU}(\pm\psi^{\ell_0}_{i,j})}{h_{\bar{W}}(\pm\psi^{\ell_0}_{i,j})}, & \rho_i \neq 1, \\
      \left(\frac{1}{k}\right)
      \frac{h_{X}(\pm\psi^{\ell_0+k}_{i,j})}{h_{\bar{W}}(\pm\psi^{\ell_0}_{i,j})} +
      \left(\frac{k-1}{k}\right)
      \frac{h_{-BU}(\pm\psi^{\ell_0}_{i,j})}{h_{\bar{W}}(\pm\psi^{\ell_0}_{i,j})},&
      \rho_i = 1,
    \end{cases}
  \end{multline}
  \fi%
  then $C^{\alpha}_{k} = \emptyset$ for all $k \geq k^\star$.
\end{thm}

The final task is to obtain the smallest
$\bar{\alpha}_k^c(\lambda_i)$ over all $k \in \mbb{N}_+$, sufficient to ensure emptiness of
$C_\infty$. This is not a simple task, for as $\psi_{i,j}^{\ell_0+k}$ processes around the set $X$ the values of $h_X(+\psi_{i,j}^{\ell_0+k})$ and $h_X(-\psi_{i,j}^{\ell_0+k})$ may increase or decrease; thus, monotonic descent of $\bar{\alpha}_k^c(\lambda_i)$ with $k$ does not follow. Nonetheless, the following and final result characterizes a condition under which the smallest value is given by considering the asymptotic behaviour of the bounds in Theorem~\ref{thm:comp1}.

\begin{thm}\label{thm:comp2}
  Consider the $ith$ Jordan block $J_i$ and suppose that the
  associated eigenvalue $\lambda_i$ is complex, with $\theta_i$ satisfying~Assumption~\ref{assump:periodic}. If, for $j=1,2$,
  \begin{equation}
    \min_{\ell \in \{0,\dots,M_i-1\}} \rho_i h_X(\pm\psi^{\ell}_{i,j}) > \max_{\ell \in \{0,\dots,M_i-1\}} h_{-BU}(\pm\psi^{\ell}_{i,j}) 
    \end{equation}
    then
    \ifCLASSOPTIONonecolumn
\begin{equation}
    \inf_{k \in \mbb{N}_+} \bar{\alpha}^c_k(\lambda_i) = \min_{j \in \{1,2\}}\min_{\ell_0, \ell \in \{0,\dots,M_i-1\}} \max \{ {\alpha}^{c,\ell}_\infty(+\psi^{\ell_0}_{i,j}), {\alpha}^{c,\ell}_\infty(-\psi^{\ell_0}_{i,j})  \},
  \end{equation}
    \else
  \begin{multline}
    \inf_{k \in \mbb{N}_+} \bar{\alpha}^c_k(\lambda_i) = \\\min_{j \in \{1,2\}}\min_{\ell_0, \ell \in \{0,\dots,M_i-1\}} \max \{ {\alpha}^{c,\ell}_\infty(+\psi^{\ell_0}_{i,j}), {\alpha}^{c,\ell}_\infty(-\psi^{\ell_0}_{i,j})  \},
  \end{multline}
  \fi%
  where%
  \begin{equation}\label{eq:infbound2}%
    {\alpha}_{\infty}^{c,\ell}(\psi^{\ell_0}_{i,j}) \coloneqq
    \begin{cases}
      \textstyle\left(\frac{1}{\rho_i}\right) \frac{h_{-BU}(\psi^{\ell_0}_{i,j})}{h_{\bar{W}}(\psi^{\ell_0}_{i,j})}, & \rho_i \geq 1, \\
      \textstyle(1-\rho_i)
      \frac{h_X(\psi^{\ell}_{i,j})}{h_{\bar{W}}(\psi^{\ell_0}_{i,j})} +
      \frac{h_{-BU}(\psi^{\ell_0}_{i,j})}{h_{\bar{W}}(\psi^{\ell_0}_{i,j})}, & \rho_i < 1.
    \end{cases}
  \end{equation}
 Consequently, if $\alpha > \inf_{k \in \mbb{N}_+} \bar{\alpha}^c_\infty(\lambda_i)$ then $C^{\alpha}_\infty = \emptyset$.
\end{thm}

Similar to how Theorems~\ref{thm:neg} and~\ref{thm:neginf} generalized
Theorems~\ref{thm:main} and~\ref{thm:inf} to reflect and capture the
alternating state behaviour associated with negative--real eigenvalue,
Theorems~\ref{thm:comp1} and \ref{thm:comp2} generalize our results to
capture the case where the evolution of the sets $A^\ell(-BU)$ and
$A^\ell \bar{W}$ is characterized by a periodic rotation and not just
a simple or alternating growth/decay. In fact, the expressions in
Theorems~\ref{thm:comp1} and \ref{thm:comp2} reduce to the earlier ones
by considering a complex eigenvalue pair with $\theta_i = 0$
(positive--real) or $\theta_i = \pi$ (negative--real).

Similar to the earlier results, Theorems~\ref{thm:comp1} and
\ref{thm:comp2} demarcate the stable and unstable system cases. Once again,
it is clear that emptiness of $C^\alpha_\infty$ depends only on the
relative sizes of the input and disturbance sets in the unstable case,
but also on the size and shape of the state set $X$ when the system is
stable. The derivation of the complex case result requires proper
consideration of the behaviour induced by a complex eigenvalue, namely
by interrogating the condition~\eqref{eq:supc} in rotating, rather
than fixed, directions.

        \section{Illustrative examples and applications}
        \label{sec:6}

    \subsection{Unstable, distinct real eigenvalues}

    Consider first
    \begin{equation}\label{eq:unstable_example}
x_{k+1} = \begin{bmatrix} \lambda_1 & 1 \\ 0 & \lambda_2 \end{bmatrix} x_k + \begin{bmatrix} 0.1 \\ 1 \end{bmatrix} u_k + w_k,
      \end{equation}
      with $\lambda_1 = 1.2$, $\lambda_2 = -1.5$; the system has two
      distinct unstable real eigenvalues. There are two linearly independent
      eigenvectors of $A^\top$, $\varphi_{1,1}$ and $\varphi_{2,1}$.
      The state, input, and basic disturbance sets are
      \begin{align}
        X &= \{ x = (x_1,x_2) : \abs{x_1}\leq 5, \abs{x_2}\leq 2\}, \\
        U &= \{ u : -0.5 \leq u \leq 1\},\\
        \bar{W} &= \{ w : \norm{w}_\infty \leq 1\}.
        \end{align}
      The values of $h_X(z)$, $h_{-BU}(z)$ and $h_{\bar{W}}(z)$ for $z = \varphi_{1,1}, \varphi_{2,1}$ are given in Table~\ref{tab:h}.
    
    \begin{table}[b]
      \centering\footnotesize
      \caption{Table of support function values, reported to 3 d.p. Since $X$ and $\bar{W}$ are symmetrical, only values in the positive directions $z=+\varphi_{i,j}$ are reported.}\label{tab:h}
      \begin{tabular}{crrrr}
        \toprule
        $z$ & $h_X(z)$ & $h_{-BU}(z)$ & $h_{-BU}(-z) $& $h_{\bar{W}}(z)$ \\
        \midrule
        \multicolumn{4}{c}{Unstable system~\eqref{eq:unstable_example}: $\lambda_1=1.2,\lambda_2=-1.5$}\\
        $\varphi_{1,1} = \begin{bmatrix} 0.94 & 0.35 \end{bmatrix}^\top$ & $5.383$ & $0.221$ & $0.441$ & $1.285$ \\
        $\varphi_{2,1} = \begin{bmatrix} 0 & 1 \end{bmatrix}^\top$ & $2.000$ & $0.500$ & $1.000$  & $1.000$ \\
        \midrule
        \multicolumn{4}{c}{Double integrator~\eqref{eq:doubleint_example}: $\lambda_1=1,\lambda_2=1$}\\
        $\varphi_{1,1} = \begin{bmatrix} 0 & 1 \end{bmatrix}^\top$ & $5.000$ & $1.000$ & $1.000$ & $1.000$ \\
        $\varphi_{1,2} = \begin{bmatrix} 1 & 0 \end{bmatrix}^\top$ & $5.000$ & $0.500$ & $0.500$ & $0.500$ \\
        \bottomrule
        \end{tabular}
      \end{table}

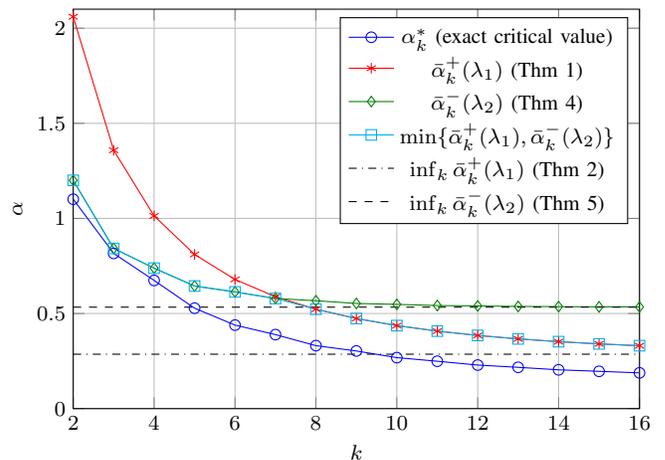
\begin{figure}[t]
  \centering\footnotesize
%
%
\definecolor{mycolor1}{rgb}{1.00000,0.00000,1.00000}%
\begin{tikzpicture}

\begin{axis}[%
width=0.85\linewidth,
height=0.6\linewidth,,
scale only axis,
xmin=2,
xmax=16,
ymin=0,
ymax=2.1,
xmajorgrids,
ymajorgrids,
ylabel = {$\alpha$},
xlabel = {$k$},
]
\addplot [color=blue, mark=o, mark options={solid, blue}]
  table[row sep=crcr]{%
2	1.101\\
3	0.816\\
4	0.674\\
5	0.528\\
6	0.439\\
7	0.389\\
8	0.331\\
9	0.303\\
10	0.268\\
11	0.249\\
12	0.229\\
13	0.217\\
14	0.204\\
15	0.196\\
16	0.188\\
};
\addlegendentry{$\alpha_k^*$ (exact critical value)}

\addplot [color=red, mark=asterisk, mark options={solid, red}]
  table[row sep=crcr]{%
2	2.06019656019656\\
3	1.35833085833086\\
4	1.01315100495428\\
5	0.810540598650683\\
6	0.679105985557598\\
7	0.588233478286548\\
8	0.522603894221307\\
9	0.473697538464958\\
10	0.43639625603152\\
11	0.407438908256094\\
12	0.384649063464536\\
13	0.366519058282774\\
14	0.351972217792277\\
15	0.340220075852263\\
16	0.330673022552882\\
};
\addlegendentry{$\bar{\alpha}^+_k(\lambda_{1})$ (Thm~\ref{thm:main})}

\addplot [color=green!50!black, mark=diamond, mark options={solid, green!50!black}]
  table[row sep=crcr]{%
2	1.2\\
3	0.842105263157895\\
4	0.738461538461539\\
5	0.644549763033175\\
6	0.613533834586466\\
7	0.578921806702283\\
8	0.567168913560666\\
9	0.552918470606645\\
10	0.548039638087031\\
11	0.541910576302549\\
12	0.539806009348719\\
13	0.537120830498868\\
14	0.536197718072792\\
15	0.535011849214373\\
16	0.534603966828741\\
};

\addlegendentry{$\bar{\alpha}^-_k(\lambda_{2})$ (Thm~\ref{thm:neg})}

\addplot [color=cyan, solid, mark=square, mark options={solid, cyan}]
  table[row sep=crcr]{%
2	1.2\\
3	0.842105263157895\\
4	0.738461538461539\\
5	0.644549763033175\\
6	0.613533834586466\\
7	0.578921806702283\\
8	0.522603894221307\\
9	0.473697538464958\\
10	0.43639625603152\\
11	0.407438908256094\\
12	0.384649063464536\\
13	0.366519058282774\\
14	0.351972217792277\\
15	0.340220075852263\\
16	0.330673022552882\\
};

\addlegendentry{$\min\{\bar{\alpha}^+_k(\lambda_{1}),\bar{\alpha}^-_k(\lambda_{2})\}$}

\addplot [color=black, dashdotted]
  table[row sep=crcr]{%
2	0.286036036036036\\
3	0.286036036036036\\
4	0.286036036036036\\
5	0.286036036036036\\
6	0.286036036036036\\
7	0.286036036036036\\
8	0.286036036036036\\
9	0.286036036036036\\
10	0.286036036036036\\
11	0.286036036036036\\
12	0.286036036036036\\
13	0.286036036036036\\
14	0.286036036036036\\
15	0.286036036036036\\
16	0.286036036036036\\
};
\addlegendentry{$\inf_{k} \bar{\alpha}^+_k(\lambda_{1})$ (Thm~\ref{thm:inf})}

\addplot [color=black, dashed]
  table[row sep=crcr]{%
2	0.533333333333333\\
3	0.533333333333333\\
4	0.533333333333333\\
5	0.533333333333333\\
6	0.533333333333333\\
7	0.533333333333333\\
8	0.533333333333333\\
9	0.533333333333333\\
10	0.533333333333333\\
11	0.533333333333333\\
12	0.533333333333333\\
13	0.533333333333333\\
14	0.533333333333333\\
15	0.533333333333333\\
16	0.533333333333333\\
};
\addlegendentry{$\inf_k \bar{\alpha}^-_k(\lambda_{2})$ (Thm~\ref{thm:neginf})}
\end{axis}
\end{tikzpicture}%
    \caption{Comparison of the critical scaling factors obtained from Theorems~\ref{thm:main} and~\ref{thm:neg} with the exact critical scaling factor $\alpha^*_{k}$ for the unstable system with $\lambda_1=1.2, \lambda_2 = -1.5$.}
    \label{fig:2}
  \end{figure}

      Theorems~\ref{thm:main} and~\ref{thm:neg} give the respective
      relevant bounds on the critical $\alpha$ for emptiness of the
      sets $C^\alpha_k$. Fig.~\ref{fig:2} compares the values of
      $\bar{\alpha}_k^+(\lambda_{1})$ (from Theorem~\ref{thm:main})
      and $\bar{\alpha}_k^-(\lambda_{2})$ (from
      Theorem~\ref{thm:neg}); a value of $\alpha$ sufficiently large
      to cause non-existence of a robust control policy is obtained by
      taking, for each $k\in\mbb{N}_+$ the minimum of the two
      values. Also shown for comparison is the exact critical scaling
      factor $\alpha_{k}^*$ for different $k \geq 2$. These were
      found by exhaustive search, for each $k \in \mbb{N}_+$ searching
      over values of $\alpha > 0$ until the smallest value is found
      for which the reachability recursion~\eqref{eq:Crecursion}
      results in $C^{\alpha}_{k} = \emptyset$; this means that
      $C^\alpha_k \neq \emptyset$ for all $\alpha <
      \alpha^*_k$.

  \subsection{Double integrator with input disturbance}

  Now we consider 
  \begin{equation}\label{eq:doubleint_example}
x_{k+1} = \begin{bmatrix} \lambda_1 & 1 \\ 0 & \lambda_2 \end{bmatrix} x_k + \begin{bmatrix} 0.5 \\ 1 \end{bmatrix} u_k + w_k,
      \end{equation}
      with $\lambda_1 = \lambda_2 = 1$ so the system has one Jordan
      block of dimension $n_i=2$; the eigenvector $\varphi_{1,1}$ and
      generalized eigenvector $\varphi_{1,2}$ are given in
      Table~\ref{tab:h}. The state, input and basic disturbance sets
      are
      \begin{align}
        X &= \{ x  : \norm{x}_\infty \leq 5\}, \\
        U &= \{ u : \abs{u} \leq 1\},\\
        \bar{W} &= \{ w = Bu : u \in U\},
      \end{align}
      so that the double integrator has an input disturbance. The values of $h_X(z)$, $h_{-BU}(z)$ and $h_{\bar{W}}(z)$ for $z = \varphi_{1,1}, \varphi_{1,2}$ are given in Table~\ref{tab:h}. 

  \begin{figure}[t]
  \centering\footnotesize
%
%
\definecolor{mycolor1}{rgb}{1.00000,0.00000,1.00000}%
\begin{tikzpicture}

\begin{axis}[%
width=0.85\linewidth,
height=0.6\linewidth,,
scale only axis,
xmin=2,
xmax=14,
ymin=0.5,
ymax=5.6,
xmajorgrids,
ymajorgrids,
ylabel = {$\alpha$},
xlabel = {$k$},
]
\addplot [color=blue, mark=o, mark options={solid, blue}]
  table[row sep=crcr]{%
2	2.76\\
3	1.56\\
4	1.19\\
5	1.05\\
6	0.98\\
7	0.94\\
8	0.93\\
9	0.92\\
10	0.92\\
11	0.91\\
12	0.91\\
13	0.91\\
14	0.91\\
};
\addlegendentry{$\alpha_k^*$ (exact critical value)};

\addplot [color=red, mark=asterisk, mark options={solid, red}]
  table[row sep=crcr]{%
2	3\\
3	2.33333333333333\\
4	2\\
5	1.8\\
6	1.66666666666667\\
7	1.57142857142857\\
8	1.5\\
9	1.44444444444444\\
10	1.4\\
11	1.36363636363636\\
12	1.33333333333333\\
13	1.30769230769231\\
14	1.28571428571429\\
};
\addlegendentry{$\bar{\alpha}^+_k(\lambda_{1})$ (Thm~\ref{thm:main})};

\addplot [color=green!50!black, mark=diamond, mark options={solid, green!50!black}]
  table[row sep=crcr]{%
2	5.5\\
3	2.8\\
4	1.9\\
5	1.52941176470588\\
6	1.34615384615385\\
7	1.24324324324324\\
8	1.18\\
9	1.13846153846154\\
10	1.10975609756098\\
11	1.08910891089109\\
12	1.07377049180328\\
13	1.06206896551724\\
14	1.05294117647059\\
};
\addlegendentry{$\bar{\beta}^+_k(\lambda_{1})$ (Thm~\ref{thm:3})};

\addplot [color=cyan, solid, mark=square, mark options={solid, cyan}]
  table[row sep=crcr]{%
2	3\\
3	2.33333333333333\\
4	1.9\\
5	1.52941176470588\\
6	1.34615384615385\\
7	1.24324324324324\\
8	1.18\\
9	1.13846153846154\\
10	1.10975609756098\\
11	1.08910891089109\\
12	1.07377049180328\\
13	1.06206896551724\\
14	1.05294117647059\\
};

\addlegendentry{$\min\{\bar{\alpha}^+_k(\lambda_{1}),\bar{\beta}^-_k(\lambda_{1})\}$}

\addplot [color=black, dashdotted]
  table[row sep=crcr]{%
2	1\\
16	1\\
};
\addlegendentry{$\inf_{k} \bar{\alpha}^+_k(\lambda_{1})$ (Thm~\ref{thm:inf})}
\end{axis}
\end{tikzpicture}%
    \caption{Comparison of the critical scaling factors obtained from Theorem~\ref{thm:main}(the eigenvector associated with $\lambda_1=1$) and~\ref{thm:3} (the generalized eigenvector associated with $\lambda_1=1$) with the exact critical scaling factor $\alpha^*_{k}$ for the double integrator.}
    \label{fig:double}
  \end{figure}
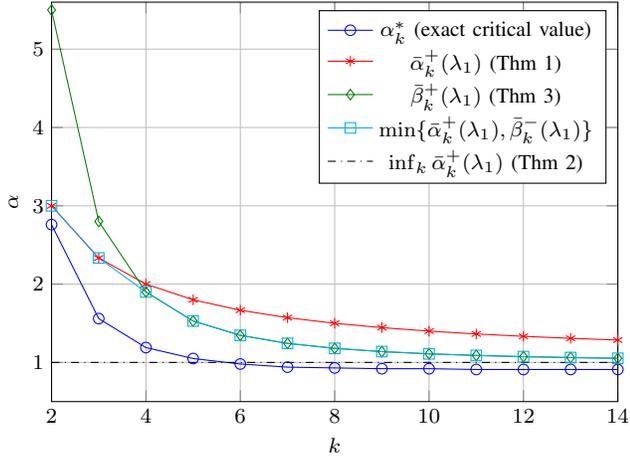

  Fig.~\ref{fig:double} shows the results. The bound from
  Theorem~\ref{thm:3}, using second-order information, is initially
  the more conservative of the two but is then tight for a range of
  $k$. The exact critical scaling factor required to make
  $C_\infty^\alpha$ empty is slightly lower than that predicted
  by~Theorem~\ref{thm:3}.

  \subsection{Adversarial attack strategies}
\label{sec:attacks}

The obtained results may be exploited to design adversarial attack strategies in a more tractable way than otherwise possible. We present this for the positive real eigenvalue case depicted in Theorem~\ref{thm:main}, but it may be extended to other cases without too much difficulty.

Consider that, given the system~\eqref{eq:dyn} and the information pattern implied by Assumption~\ref{assump:basic}, an attacker seeks to drive the state $x_k$ outside of the set $X$ despite the best actions of the controller; that is, given $x_0\in X$, to find a sequence $\{w_k\}$ such that
\begin{equation}
  (\exists k^* \in \mbb{N})(\forall \mb{u}_{k^*} \in \mc{U}_{k^*}) \ x_{k^*} \not\in X.
\end{equation}
Such a sequence design problem may be expressed as
\begin{equation}\label{eq:attackprob}
\max_{\mb{w}_N \in \alpha \bar{\mc{W}}_N} \min_{\mb{u}_N \in \mc{U}_N} \sum_{k=1}^{N} \mathbbm{1}_{X}(x_{k}) 
  \end{equation}
  where the use of the indicator function means this problem serves as a test for system safety, yielding a binary outcome: $0$ if the controller can guarantee safety, but $\infty$ if the attacker can force a state constraint violation; however, this bilevel optimization problem is intractable in general. Nonetheless, supposing that the attacker has knowledge of an $\alpha$ satisfying~\eqref{eq:alpha111} and also $\varphi_{i,1}$, the corresponding real eigenvector of $A^\top$, the problem is considerably simplified by considering the projected state
  \begin{equation}
    \xi_{k} \coloneqq \varphi_{i,1}^\top x_k \in \mbb{R}
  \end{equation}
  so that the system evolves in the direction $\varphi_{i,1}$ according to the scalar dynamics
  \begin{equation}\label{eq:ximodel}
    \begin{split}
      \xi_{k+1} &= \varphi^\top_{i,1} A x_k + \varphi_{i,1}^\top B u_k + \varphi_{i,1}^\top w_k \\
      &= \lambda_i \xi_{k} + \upsilon_k + \omega_k,
      \end{split}
    \end{equation}
    where $\upsilon_k \coloneqq \varphi_{i,1}^\top B u_k \in \mbb{R}$, $\omega_k \coloneqq \varphi_{i,1}^\top w_k \in \mbb{R}$. Problem~\eqref{eq:attackprob} is then
    \begin{equation}\label{eq:attackprob_scalar}
\max_{\{\omega_0,\dots,\omega_{N-1}\} \in \alpha \bar{\Omega}} \min_{\{\upsilon_0,\dots,\upsilon_N\} \in \Upsilon} \sum_{k=1}^N  \mathbbm{1}_{\Xi}(\xi_{k}), 
\end{equation}
subject to~\eqref{eq:ximodel},
  which is considerably simpler in view of the scalar dynamics and the simple bounds on variables defined by support function values
  \begin{align}
    \bar{\Omega} &\coloneqq \varphi_{i,1}^\top \bar{W} = \left[ -h_{\bar{W}}(-\varphi_{i,1}), +h_{\bar{W}}(+\varphi_{i,1})\right],\\
    \Upsilon &\coloneqq \varphi_{i,1}^\top BU = \left[ -h_{BU}(-\varphi_{i,1}), +h_{BU}(+\varphi_{i,1})\right],\\
    \Xi &\coloneqq \varphi_{i,1}^\top X = \left[ -h_{X}(-\varphi_{i,1}), +h_{X}(+\varphi_{i,1})\right].
    \end{align}

    Then, starting from $\xi_0 = \varphi_{i,1}^\top x_0$, a heuristic to solve~\eqref{eq:attackprob} is repeatedly apply the most positive or most negative disturbance, depending on the sign of $\xi_k$:
    \begin{equation}
      \omega_k(\xi_k) = \begin{dcases} +\alpha h_{\bar{W}}(+\varphi_{i,1}), & \xi_k > 0, \\
        -\alpha h_{\bar{W}}(-\varphi_{i,1}), & \xi_k < 0, \end{dcases}
      \end{equation}
      assuming that the defender controlling the system selects the input that pushes the state in the opposite direction, back into $X$:
      \begin{equation}
      \upsilon_k(\xi_k) = \begin{dcases}  -h_{BU}(-\varphi_{i,1}), & \xi_k > 0, \\
        + h_{BU}(+\varphi_{i,1}), & \xi_k < 0. \end{dcases}
    \end{equation}
    This heuristic, which can be interpreted as a greedy attack policy exploiting the scalar nature of $\xi_k$, maximizes the rate of growth of $\xi_k$ in the sign-consistent direction. Because of Theorem~\ref{thm:main}, it is guaranteed that the state $x_k$ will, in finite time, leave the set $X$ in the direction $+\varphi_{i,1}$ or $-\varphi_{i,1}$.

    \section{Conclusions and future work}
    \label{sec:8}

    This paper has derived sufficient conditions on the size of a
    scaled disturbance set in order that there does not exist a robust
    control policy capable of keeping the state of a linear
    system within a given constraint set for all realizations of
    the disturbance. The conditions are algebraic and closed-form in
    nature, and expose directly the dependence of the critical scaling
    factor on the shapes and sizes of state, input and disturbance
    sets, plus the spectral properties of the system. We considered a
    general linear system with all possible arrangements and
    multiplicities of eigenvalues, and developed different conditions
    for positive real, negative real, and complex
    eigenvalues. Numerical investigations illustrated that the
    obtained bounds are sometimes tight and in other cases
    conservative, which should be investigated in further work. Future
    directions for research include the converse problem: what is a
    sufficiently small disturbance set (or sufficiently large input
    set) in order to guarantee the existence of a robust control
    policy.

\appendix

\subsection{Proof of Proposition~\ref{prop:Tdef}}\label{proof:Tdef}
We have ${C}_0 = X = A^{-0}T_0$ and
  \begin{equation}
    \begin{split}
      {C}_1 &= A^{-1}\left( [X \ominus W] \oplus (-BU) \right) \cap {X} \\
    &= A^{-1} \left( [T_0 \ominus W] \oplus (-BU) \right)\cap A^{-0} T_0\\
    &= A^{-1} T_1 \cap A^{-0} T_0 .
    \end{split}
    \end{equation}
    Now ${C}_2 = A^{-1} \left([{C}_1 \ominus W] \oplus (-BU) \right) \cap X$, so 
    \ifCLASSOPTIONonecolumn
\begin{equation}\label{eq:C2}
\begin{split}
  C_2 &= A^{-1} \left([ (A^{-1} T_1 \cap A^{-0} T_0) \ominus {W}] \oplus {(-BU)} \right) \cap X\\
  &= A^{-1} \left([ (A^{-1} T_1 \ominus {W}) \cap (A^{-0} {T}_0 \ominus {W})] \oplus {(-BU)} \right) \cap {X}\\
  &\subseteq A^{-1} \left([ (A^{-1} {T}_1 \ominus {W})  \oplus  {(-BU)}] \cap [(A^{-0} {T}_0 \ominus {W})  \oplus  {(-BU)}]\right) \cap  {X}
\end{split}
\end{equation}
    \else
    \begin{equation}\label{eq:C2}
\begin{split}
  C_2 &= A^{-1} \left([ (A^{-1} T_1 \cap A^{-0} T_0) \ominus {W}] \oplus {(-BU)} \right) \cap X\\
  &= A^{-1} \left([ (A^{-1} T_1 \ominus {W}) \cap (A^{-0} {T}_0 \ominus {W})] \oplus {(-BU)} \right) \cap {X}\\
  &\subseteq A^{-1} \left([ (A^{-1} {T}_1 \ominus {W})  \oplus  {(-BU)}] \right. \\ &\quad \left. \cap [(A^{-0} {T}_0 \ominus {W})  \oplus  {(-BU)}]\right) \cap  {X}
\end{split}
\end{equation}
\fi%
where the second and third lines, respectively, follow from the facts that the Minkowski difference is distributive over intersection~\cite{KG98}:
\begin{equation}
(X \cap Y) \ominus Z = (X \ominus Z) \cap (Y \ominus Z),
\end{equation}
but the sum is only subdistributive over intersection~\cite{Schneider93}:
\begin{equation}
(X \cap Y) \oplus Z \subseteq (X \oplus Z) \cap (Y \oplus Z).
\end{equation}
Using the fact $A^{-1}(X \cap Y) = A^{-1} X \cap A^{-1} Y$, the right-hand side of~\eqref{eq:C2} is equal to
\ifCLASSOPTIONonecolumn
\begin{equation}
 A^{-2}( {T}_1 \ominus A{W} \oplus A{(-BU)} ) \cap A^{-1}( {T}_0 \ominus {W} \oplus {(-BU)}) \cap {T}_0
  = A^{-2} {T}_2 \cap A^{-1} {T}_1 \cap A^{-0} {T}_0.
\end{equation}
\else
\begin{multline}
 A^{-2}( {T}_1 \ominus A{W} \oplus A{(-BU)} ) \cap A^{-1}( {T}_0 \ominus {W} \oplus {(-BU)}) \cap {T}_0\\
  = A^{-2} {T}_2 \cap A^{-1} {T}_1 \cap A^{-0} {T}_0.
\end{multline}
\fi%
Thus, $C_2 \subseteq \bigcap_{\ell=0}^2 A^{-\ell} T_\ell$. Similar arguments establish the same for $C_k $ for all $k > 2$.  \hfill\IEEEQED

\subsection{Proof of Proposition~\ref{prop:Cexist}}\label{proof:Cexist}

By Proposition~\ref{prop:Tdef}, $C_k \subseteq \bigcap_{\ell=0}^k A^{-\ell} T_\ell$.
            It follows that, if $T_{k^\star} = \emptyset$ for some
            $k^\star$, then $C_k = \emptyset$ for all
            $k \geq k^\star$. \hfill \IEEEQED

            \subsection{Proof of Proposition~\ref{prop:Texist}}\label{proof:Texist}

        Consider $T_{k+1} = (T_k \ominus A^k W ) \oplus A^kB(-U)$. Applying this recursively from $T_0 = X$, we obtain
        \ifCLASSOPTIONonecolumn
\begin{equation}
      T_{k} = X \ominus A^0W \oplus A^0B(-U)  \ominus A^1W \oplus A^1B(-U)  \dots  \ominus A^{k-1} W \oplus A^{k-1} B(-U)
    \end{equation}
        \else
    \begin{multline}
      T_{k} = X \ominus A^0W \oplus A^0B(-U) \\ \ominus A^1W \oplus A^1B(-U) \\ \vdots \\ \ominus A^{k-1} W \oplus A^{k-1} B(-U)
    \end{multline}
    \fi
      For non-empty sets $A$, $B$ and $C$, we have $(A \ominus B) \oplus C \subseteq (A \oplus C) \ominus B$. Therefore, for all $k \geq 0$, 
      \begin{equation}
        T_{k-1} \subseteq X \oplus \left[\bigoplus_{\ell=0}^{k-2} A^\ell B(-U)\right] \ominus \left[\bigoplus_{\ell=0}^{k-2} A^\ell W \right]
        \end{equation}
        Consider how $T_k$ is determined from $T_{k-1}$: first, $A^{k-1}W$ is subtracted. It follows that
        \begin{equation}
          T_{k^\star} = \emptyset \iff T_{k^\star-1} \ominus A^{{k^\star}-1}W = \emptyset \ \text{for some} \ k^\star \geq 1.
        \end{equation}
          The latter is true if (but not only if), for $k^\star \geq 1$,
          \begin{equation}
            X \oplus \left[\bigoplus_{\ell=0}^{k^\star-2} A^\ell B(-U)\right] \ominus \left[\bigoplus_{\ell=0}^{k^\star-2} A^\ell W \right] \ominus A^{k^\star-1}W = \emptyset,
            \end{equation}
            rearrangement of which gives the stated condition. \hfill \IEEEQED

\ifextended

\subsection{Proof of Lemma~\ref{lem:hbound}}

Consider, for any $i \in \mc{M}\coloneqq \{1,\dots,M\}$,
\begin{multline}
h_Y(z_i) = h_Y\bigg(\sum_{j \in \mc{M}} z_j - \sum_{j \in \mc{M} \setminus \{i\}} z_j \bigg) \\
\leq h_Y\bigg(\sum_{j \in \mc{M}} z_j\bigg) + h_Y\bigg(-\sum_{j \in \mc{M} \setminus \{i\}} z_j\bigg).
\end{multline}
Then, because $h_Y(-z) = h_{-Y}(z)$ for a C-set $Y$ and any $z\in\mbb{R}^n$, we have, for any $i \in \mc{M}$,
\begin{multline}
  h_Y(z_i) \leq h_Y\bigg(\sum_{j\in\mc{M}} z_j\bigg) + h_{-Y}\bigg(\sum_{j\in\mc{M}\setminus\{i\}} z_j\bigg) \\
  \leq h_Y\bigg(\sum_{j\in\mc{M}} z_j\bigg) + \sum_{j \in \mc{M}\setminus\{i\}} h_{-Y}(z_j).
\end{multline}
where we have use the subadditivity property of $h_Y$~\cite{Schneider93}. Rearranging,
\begin{equation}
  h_Y(z_i) - \sum_{j \in \mc{M} \setminus \{i\}} h_{-Y}(z_j) \leq h_Y\bigg(\sum_{j \in \mc{M}} z_j\bigg).
\end{equation}
The greatest lower bound is obtained by taking the maximum of the left-hand side over $i \in \mc{M}$. \hfill\IEEEQED

\subsection{Proof of Lemma~\ref{lem:f}}

Recall $f_i(\ell) \coloneqq \lambda_i^\ell h_{\bar{W}}(\varphi_{i,2}) - \ell \lambda_i^{\ell-1} h_{\bar{W}}(\varphi_{i,1})$. For (i), if $\ell=0$ then $f_i(\ell) = \lambda_i^0 h_{\bar{W}}(\varphi_{i,2}) - 0 \lambda_i^{-1} h_{\bar{W}}(\varphi_{i,1}) = h_{\bar{W}}(\varphi_{i,2})$. For (ii), consider $f_i(\ell) = 0$. Then
\begin{equation}
\lambda_i^\ell h_{\bar{W}}(\varphi_{i,2}) = \ell \lambda_i^{\ell-1} h_{\bar{W}}(\varphi_{i,1})  \iff \ell = \frac{\lambda_i^\ell}{\lambda_i^{\ell-1}}\frac{h_{\bar{W}}(\varphi_{i,2})}{h_{\bar{W}}(\varphi_{i,1})},
\end{equation}
which is well defined since $h_{\bar{W}}(z) > 0$ for all $z\in\mbb{R}^n$ and $\lambda_i > 0$. The solution for $\ell$ simplifies to
\begin{equation}
\ell = \lambda_i\frac{h_{\bar{W}}(\varphi_{i,2})}{h_{\bar{W}}(\varphi_{i,1})} \eqqcolon \ell^*_i.
\end{equation}
For (iii), consider $0 \leq \ell < \ell_i^*$. Then
\begin{equation}
0 \leq \ell < \lambda_i\frac{h_{\bar{W}}(\varphi_{i,2})}{h_{\bar{W}}(\varphi_{i,1})} = \frac{\lambda_i^\ell}{\lambda_i^{\ell-1}}\frac{h_{\bar{W}}(\varphi_{i,2})}{h_{\bar{W}}(\varphi_{i,1})}.
  \end{equation}
  Therefore,
  \begin{equation}
    \ell \lambda_i^{\ell-1} h_{\bar{W}}(\varphi_{i,1}) < \lambda_i^\ell h_{\bar{W}}(\varphi_{i,2}) \implies f_i(\ell) > 0.
    \end{equation}
    A similar argument leads to $f_i(\ell) < 0$ for $\ell > \ell^*_i$, establishing (iv). Part (v) follows from (iii) and (iv) since $+f_i(\ell) > 0 > -f_i(\ell)$ for $0\leq \ell < \ell^*_i$, with the opposite inequalities for $\ell > \ell_i^*$. Finally, (vi) follows directly from the fact that $\max\{ f_i(\ell), -f_i(\ell) \}> 0$ for all $\ell \geq 0$. \hfill\IEEEQED

    \subsection{Proof of Theorem~\ref{thm:3}}\label{proof:3}

    The proof is given for the positive direction $z=+\varphi_{i,2}$ but is easily extended to the opposite direction via an appropriate change of sign in support function arguments. Each analysis yields a condition for negativity of $h_{S_k^\alpha}(z)$; given that $h_{S_k^\alpha}(+\varphi_{i,2}) < 0$ and $h_{S_k^\alpha}(+\varphi_{i,2}) < 0$ is the sufficient condition for emptiness of $S_k^\alpha$, the sufficient value of $\alpha$ is obtained by taking the maximum of the two conditions.

    Inserting~\eqref{eq:tbound} and~\eqref{eq:bbound} into the right-hand side~\eqref{eq:35} of the inequality~\eqref{eq:34}, we obtain
\begin{equation}\label{eq:Sa2}
\begin{split}
  (\forall k \in \mbb{N}) \ h_{S^\alpha_k}(\varphi_{i,2}) &\leq h_X(\varphi_{i,2}) 
  + \bigg(\sum_{\ell=0}^{k-2} \lambda_{i}^\ell \bigg) h_{-BU}(\varphi_{i,2}) \\
  &\quad + \bigg(\sum_{\ell=0}^{k-2} \ell \lambda_{i}^{\ell-1} \bigg) h_{-BU}(\varphi_{i,1}) \\
  &\quad - \alpha \sum_{\ell=0}^{k-1} \max \{ f_i(\ell), -f_i(\ell) \}.
\end{split}
\end{equation}
      Given that $\lambda_i > 0$, $h_X(\varphi_{i,2}) > 0$, $h_{-BU}(\cdot) \geq 0$ and, by Lemma~\ref{lem:f}, $\sum_{\ell=0}^{k-1} \max \{ f_i(\ell), -f_i(\ell)\} > 0$ for all $k\geq 1$, the right-hand side of~\eqref{eq:Sa2} is negative for all pairs $(\alpha,k)$ satisfying
      \begin{equation}
        \begin{dcases}\begin{aligned}
        \alpha &> 0, k \geq 1, \\
        \alpha &>  \frac{h_X(\varphi_{i,2})  + s_{0,k-2} h_{-BU}(\varphi_{i,2})  + s^\prime_{0,k-2}h_{-BU}(\varphi_{i,1})}{\sum_{\ell=0}^{k-1} \max \{ f_i(\ell), -f_i(\ell)\} }, \end{aligned}\end{dcases}\label{eq:80}
      \end{equation}
      where we have used the shorthand in~\eqref{eq:geo1} and~\eqref{eq:geo2} for the partial sums in the numerator.

Because of Lemma~\ref{lem:f}, 
  \begin{multline}\label{eq:fwithk}
    \sum_{\ell=0}^{k-1} \max\{ f_i(\ell), -f_i(\ell) \} \\
    = \begin{cases} \sum_{\ell=0}^{k-1}f_i(\ell), & 1 \leq k \leq \floor*{\ell_i^*}+1. \\ \sum_{\ell=0}^{\floor*{\ell^*_i}} f_i(\ell) + \sum_{\ell=\floor*{\ell^*_i}+1}^{k-1} -f_i(\ell), & k > \floor*{\ell_i^*}+1. \end{cases}
  \end{multline}
  Consider the simpler $k \leq \floor*{\ell_i^*}+1$ case first. Then,
  \begin{multline}\label{eq:85}
(\forall k \in \{1,\dots,\floor*{\ell_i^*}+1\}) \ \sum_{\ell=0}^{k-1} \max\{f_i(\ell),-f_i(\ell)\} \\=  s_{0,k-1} h_{\bar{W}}(\varphi_{i,2})  - s^\prime_{0,k-1}  h_{\bar{W}}(\varphi_{i,1}).
\end{multline}
Thus, the condition~\eqref{eq:80} becomes
\begin{equation}
\alpha >  \frac{h_X(\varphi_{i,2}) + s_{0,k-2} h_{-BU}(\varphi_{i,2}) 
      + s^\prime_{0,k-2} h_{-BU}(\varphi_{i,1})}{s_{0,k-1} h_{\bar{W}}(\varphi_{i,2}) - s^\prime_{0,k-1}  h_{\bar{W}}(\varphi_{i,1})}.
\end{equation}
The final form~\eqref{eq:alpha2} is reached by replacing the geometric series using~\eqref{eq:geo1}
and~\eqref{eq:geo2} with suitably adjusted indices.

The other case, where $k > \floor*{\ell_i^*}+1$, remains. From~\eqref{eq:fwithk} and the definition~\eqref{eq:f} of $f_i$,
  \begin{multline}
    (\forall k > \floor*{\ell_i^*}+1) \ \sum_{\ell=0}^{k-1} \max\{ f_i(\ell), -f_i(\ell) \} \\= \sum_{\ell=0}^{\floor*{\ell^*_i}} \lambda_i^\ell h_{\bar{W}}(\varphi_{i,2}) - \ell \lambda_i^{\ell-1} h_{\bar{W}}(\varphi_{i,1}) \\ + \sum_{\ell=\floor*{\ell^*_i}+1}^{k-1}  \ell \lambda_i^{\ell-1} h_{\bar{W}}(\varphi_{i,1}) - \lambda_i^\ell h_{\bar{W}}(\varphi_{i,2}).
  \end{multline}
  After gathering common terms,
  \begin{equation}\label{eq:87}
    \sum_{\ell=0}^{k-1} \max\{ f_i(\ell), -f_i(\ell) \} \\=     G_2 h_{\bar{W}}(\varphi_{i,2}) 
    + G_1 h_{\bar{W}}(\varphi_{i,1}),
  \end{equation}
  where
  %
  %
  \begin{align}
    G_1 &\coloneqq s^\prime_{\floor*{\ell_i^*}+1,k-1} - s^\prime_{0,\floor*{\ell_i^*}} \\
    G_2 &\coloneqq s_{0,\floor*{\ell_i^*}} - s_{\floor*{\ell_i^*}+1,k-1}.
    \end{align}
  
  Therefore, condition~\eqref{eq:80} becomes
  \begin{equation}\label{eq:alphaG1G2}
\alpha >  \frac{ h_X(\varphi_{i,2}) + s_{0,k-2} h_{-BU}(\varphi_{i,2}) 
      + s^\prime_{0,k-2} h_{-BU}(\varphi_{i,1})}{G_2 h_{\bar{W}}(\varphi_{i,2}) + G_1 h_{\bar{W}}(\varphi_{i,1})},
  \end{equation}
where, again, the geometric series in the numerator are straightforwardly replaced using~\eqref{eq:geo1} and~\eqref{eq:geo2}, but the coefficients in the denominator require a little more work.

Consider $G_2$ first. The first term is
  \begin{equation}
    s_{0,\floor*{\ell_i^*}} = \begin{cases} \frac{1 - \lambda_i^{\floor*{\ell_i^*}+1}}{1 - \lambda_i}, & \lambda_i \neq 1, \\
    \floor*{\ell_i^*}+1, & \lambda_i = 1,
    \end{cases}
    \end{equation}
    and the second term is
  \begin{equation}
    s_{\floor*{\ell_i^*}+1,k-1}  = \begin{cases}  \frac{\lambda_{i}^{\floor*{\ell_i^*}+1} - \lambda_i^{k}}{1 - \lambda_i}, & \lambda_i \neq 1, \\
    k - \floor*{\ell_i^*} - 1, & \lambda_i = 1.
    \end{cases}
  \end{equation}
  Therefore, 
  \begin{equation}\label{eq:G2exp}
    G_2 = \begin{cases} \frac{1-2\lambda_i^{\floor*{\ell_i^*}+1}  + \lambda_i^k}{1-\lambda_i}, & \lambda_i \neq 1,\\
      2(\floor*{\ell_i^*}+1) - k, &\lambda_i=1.
    \end{cases}
    \end{equation}

    Now consider $G_1$. The second term is 
    \begin{equation}
      s^\prime_{0,\floor*{\ell_i^*}}= \begin{cases} \frac{1 - (\floor*{\ell_i^*}+1)\lambda_i^{\floor*{\ell_i^*}} + \floor*{\ell_i^*}\lambda_i^{\floor*{\ell_i^*}+1}}{(1-\lambda_i)^2}, & \lambda_i \neq 1 \\
    \frac{\floor*{\ell_i^*}(\floor*{\ell_i^*}+1)}{2}, & \lambda_i = 1,
    \end{cases}
      \end{equation}
      and the first term is
        %
        \begin{multline}
          s^\prime_{\floor*{\ell_i^*}+1,k-1}=\\ \begin{cases} \frac{(\floor*{\ell_i^*}+1)\lambda_i^{\floor*{\ell_i^*}} - \floor*{\ell_i^*}\lambda_i^{\floor*{\ell_i^*}+1} + (k-1)\lambda_i^{k} - k \lambda_i^{k-1} }{(1-\lambda_i)^2}, & \lambda_i \neq 1 \\
    \frac{(k-\floor*{\ell_i^*}-1)(\floor*{\ell_i^*}+k)}{2}, & \lambda_i = 1.
    \end{cases}
      \end{multline}
      Therefore,
      \begin{multline}
        G_1 = \\\begin{cases} \frac{2(\floor*{\ell_i^*}+1)\lambda_i^{\floor*{\ell_i^*}} - 2\floor*{\ell_i^*}\lambda_i^{\floor*{\ell_i^*}+1} + (k-1)\lambda_i^{k} - k \lambda_i^{k-1} - 1}{(1-\lambda_i)^2}, & \lambda_i \neq 1 \\
    \frac{k^2- k - 2\floor*{\ell_i^*}^2 - 2\floor*{\ell_i^*}}{2}, & \lambda_i = 1.
          \end{cases}
        \end{multline}
        Inserting~$G_1$ and $G_2$ into~\eqref{eq:alphaG1G2} yields the corresponding expression in~\eqref{eq:alpha2}. \hfill\IEEEQED

\subsection{Proof of Proposition~\ref{prop:alphaseqneg}}\label{proof:alphaseqneg}

The value of ${\alpha}^-_k(\pm\varphi_{i,1})$, given in~\eqref{eq:alphaminus}, is positive for all $k\in\mbb{N}_+$ because $h_X(z), h_{-BU}(z), h_{\bar{W}} > 0$ for all $z\in\mbb{R}^n$ and all coefficients of these terms are, for all $k\in\mbb{N}_+$, positive. Consider a $k \in \mbb{N}_+$, which may be odd or even, and analyse the sequence $\{\alpha^-_k(+\varphi_{i,1})\}$. We have
\begin{multline}
{\alpha}^-_{k}(\varphi_{i,1}) = \\ \frac{h_X(\varphi_{i,1}) + a_{k-2} h_{-BU}(-\varphi_{i,1}) + b_{k-2} h_{-BU}(\varphi_{i,1})}{a_{k-1} h_{\bar{W}}(-\varphi_{i,1}) + b_{k-1} h_{\bar{W}}(\varphi_{i,1})},
\end{multline}
where
\begin{equation}
a_k \coloneqq \sum_{\substack{\ell=0\\\ell \, \text{odd}}}^{k} \abs{\lambda_i}^\ell \ \text{and} \ b_k \coloneqq \sum_{\substack{\ell=0\\\ell \, \text{even}}}^{k} \abs{\lambda_i}^\ell.
\end{equation}
To account for the compare ${\alpha}^-_{k+1}(\varphi_{i,1})$ with ${\alpha}^-_{k}(\varphi_{i,1})$, but rather ${\alpha}^-_{k+2}(\varphi_{i,1})$; if $k$ is odd (even) then the conditions we obtain are for increase (decrease) in ${\alpha}^-_{k}(\varphi_{i,1})$ over odd (even) steps. We have, therefore,
\begin{multline}
{\alpha}^-_{k+2}(\varphi_{i,1}) = \\ \frac{h_X(\varphi_{i,1}) + a_{k} h_{-BU}(-\varphi_{i,1}) + b_{k} h_{-BU}(\varphi_{i,1})}{a_{k+1} h_{\bar{W}}(-\varphi_{i,1}) + b_{k+1} h_{\bar{W}}(\varphi_{i,1})},
\end{multline}%
and wish to consider the difference ${\alpha}^-_{k+2}(\varphi_{i,1}) - {\alpha}^-_{k}(\varphi_{i,1})$.

Suppose first that $k$ is \emph{odd}; then $k+1$ and $k-1$ are even and $k-2$ is odd, and
\begin{equation}
(\forall k \, \text{odd}) \ \begin{dcases} \begin{aligned}  a_{k-1} &= a_{k-2}, & a_k &= a_{k-2} + \abs{\lambda_i}^k,\\
  b_{k} &= b_{k-1}, & b_{k-1} &= b_{k-2} + \abs{\lambda_i}^{k-1}.\end{aligned}\end{dcases}\label{eq:bs}
\end{equation}
Denote ${\alpha}^-_{k}(\varphi_{i,1}) = N_k/D_k$ and ${\alpha}^-_{k+2}(\varphi_{i,1}) = N_{k+2}/D_{k+2}$. Then
\begin{equation}\label{eq:alpk+2-k}
  {\alpha}^-_{k+2}(\varphi_{i,1}) - {\alpha}^-_{k}(\varphi_{i,1}) = \frac{N_{k+2}D_k - N_kD_{k+2}}{D_{k+2}D_{k}}.
\end{equation}
Since $D_{k+2}D_{k} > 0$ for all $k\in\mbb{N}_+$, positivity or negativity of the difference~\eqref{eq:alpk+2-k} is determined by the numerator. Exploiting~\eqref{eq:bs}, we find that
\begin{multline}
N_{k+2}D_k - N_kD_{k+2}\\=  h_{\bar{W}}(\varphi_{i,1}) \left(\abs{\lambda_i}^k h_{-BU}(-\varphi_{i,1}) +  \abs{\lambda_i}^{k-1} h_{-BU}(\varphi_{i,1})\right)  \\-h_X(\varphi_{i,1}) \left(\abs{\lambda_i}^k h_{\bar{W}}(-\varphi_{i,1}) +  \abs{\lambda_i}^{k+1} h_{\bar{W}}(\varphi_{i,1})\right).
\end{multline}
Therefore, with $k$ odd,~\eqref{eq:alpk+2-k} is strictly negative if
\begin{multline}\label{eq:hineq}
 \left( \abs{\lambda_i} h_{\bar{W}}(-\varphi_{i,1}) +  \abs{\lambda_i}^2 h_{\bar{W}}(\varphi_{i,1})\right) h_X(\varphi_{i,1}) >\\\left( \abs{\lambda_i} h_{-BU}(-\varphi_{i,1}) +  h_{-BU}(\varphi_{i,1})\right) h_{\bar{W}}(\varphi_{i,1})
  \end{multline}
  and strictly positive if the inequality in~\eqref{eq:hineq} has the opposite direction; thus $\{ {\alpha}^-_{k}(\varphi_{i,1})\}_{k \, \text{odd}}$ is positive and decreasing (increasing).

  Then, supposing that $k$ is \emph{even}, and using the facts
  \begin{equation}
(\forall k \, \text{even}) \ \begin{dcases} \begin{aligned}  a_{k} &= a_{k-1}, & a_{k-1} &= a_{k-2} + \abs{\lambda_i}^{k-1},\\
  b_{k-1} &= b_{k-2}, & b_{k} &= b_{k-2} + \abs{\lambda_i}^{k},\end{aligned}\end{dcases}\label{eq:bsnew}
\end{equation}
we find that
\begin{multline}
N_{k+2}D_k - N_kD_{k+2} \\=  h_{-BU}(-\varphi_{i,1}) \left(\abs{\lambda_i}^{k} h_{\bar{W}}(-\varphi_{i,1}) +  \abs{\lambda_i}^{k-1} h_{\bar{W}}(\varphi_{i,1})\right)  \\-h_X(\varphi_{i,1}) \left(\abs{\lambda_i}^{k+1} h_{\bar{W}}(-\varphi_{i,1}) +  \abs{\lambda_i}^{k} h_{\bar{W}}(\varphi_{i,1})\right)
\end{multline}
Therefore, with $k$ even,~\eqref{eq:alpk+2-k} is strictly negative if
\begin{multline}\label{eq:hineq2}
 \left( \abs{\lambda_i}^2 h_{\bar{W}}(-\varphi_{i,1}) +  \abs{\lambda_i} h_{\bar{W}}(\varphi_{i,1})\right) h_X(\varphi_{i,1}) >\\\left( \abs{\lambda_i} h_{\bar{W}}(-\varphi_{i,1}) +  h_{\bar{W}}(\varphi_{i,1})\right) h_{-BU}(-\varphi_{i,1})
  \end{multline}
  and strictly positive if the inequality in~\eqref{eq:hineq2} has the opposite direction; thus $\{ {\alpha}^-_{k}(\varphi_{i,1})\}_{k \, \text{even}}$ is positive and decreasing (increasing).
  
The counterpart results for $\{\alpha_{k}^-(-\varphi_{i,1})\}$ are obtained by performing an appropriate change of sign in the support function arguments in the derived conditions. \hfill \IEEEQED

\else
\fi

\subsection{Proof of Theorem~\ref{thm:neginf}}
\label{proof:neginf}

We are interested in finding
\begin{equation}\label{eq:inffind}
\inf_{k \in \mbb{N}} \max \{ \alpha_k^-(+\varphi_{i,1}), \alpha_k^-(-\varphi_{i,1}) \}, 
\end{equation}
which is the minimum of
\begin{equation}
\inf_{n \in \mbb{N}} \max \{ \alpha_{2n+1}^-(+\varphi_{i,1}), \alpha_{2n+1}^-(-\varphi_{i,1}) \}
\end{equation}
and
\begin{equation}
\inf_{n \in \mbb{N}_+} \max \{ \alpha_{2n}^-(+\varphi_{i,1}), \alpha_{2n}^-(-\varphi_{i,1}) \}.
\end{equation}
If~Proposition~\ref{prop:alphaseqneg} holds, then $\{ {\alpha}^-_{2n+1}(\pm\varphi_{i,1})\}_{n \in\mbb{N}}$ and $\{ {\alpha}^-_{2n}(\pm\varphi_{i,1})\}_{n\in\mbb{N}_+}$ are strictly decreasing, positive sequences. Thus, the greatest lower bound of each sequence coincides with its limit, and~\eqref{eq:inffind} is the minimum of
\begin{equation}
\max \left\{ \lim_{n\to\infty} \alpha_{2n+1}^-(+\varphi_{i,1}), \lim_{n \to \infty} \alpha_{2n+1}^-(-\varphi_{i,1}) \right\}
\end{equation}
and
\begin{equation}
\max \left\{ \lim_{n\to\infty} \alpha_{2n}^-(+\varphi_{i,1}), \lim_{n\to\infty} \alpha_{2n}^-(-\varphi_{i,1}) \right\},
\end{equation}
as claimed.
  
  To find the limits, consider~\eqref{eq:alphaminus} in the positive direction $+\varphi_{i,1}$. For $k$ \emph{odd}, $2\floor{(k-1)/2} = 2\ceil{(k-1)/2} = k-1$ and $2\floor{k/2} = k-1$, while $2\ceil{k/2} = k+1$. Thus, $\alpha_{k}^- (\varphi_{i,1})$ is, for $k$ odd,
  \begin{equation}
    \frac{h_X(\varphi_{i,1}) + \frac{\abs{\lambda_i} - \abs{\lambda_i}^k}{1-\abs{\lambda_i}^2}h_{-BU}(-\varphi_{i,1}) + \frac{1-\abs{\lambda_i}^{k-1}}{1-\abs{\lambda_i}^2}h_{-BU}(\varphi_{i,1})  }{\frac{\abs{\lambda_i} - \abs{\lambda_i}^k}{1-\abs{\lambda_i}^2}h_{\bar{W}}(-\varphi_{i,1}) + \frac{1 - \abs{\lambda_i}^{k+1}}{1-\abs{\lambda_i}^2}h_{\bar{W}}(\varphi_{i,1}) },
  \end{equation}
  which has the limit
  \begin{equation}\label{eq:lim11}
    \frac{(1-\abs{\lambda_i}^2) h_X(\varphi_{i,1}) + \abs{\lambda_i} h_{-BU}(-\varphi_{i,1}) + h_{-BU}(\varphi_{i,1}) }{\abs{\lambda_i} h_{\bar{W}}(-\varphi_{i,1}) + h_{\bar{W}}(\varphi_{i,1})}
  \end{equation}
  if $\abs{\lambda_i} < 1$, and
  \begin{equation}\label{eq:lim22}
   \frac{\abs{\lambda_i} h_{-BU}(-\varphi_{i,1}) + h_{-BU}(\varphi_{i,1}) }{\abs{\lambda_i} h_{\bar{W}}(-\varphi_{i,1}) + \abs{\lambda_i}^2 h_{\bar{W}}(\varphi_{i,1})},
 \end{equation}
 if $\abs{\lambda_i} \geq 1$. If~\eqref{eq:negcond1} is met, $\inf_{n \in\mbb{N}} \alpha_{2n+1}^- (\varphi_{i,1})= \lim_{n \to\infty} \alpha_{2n+1}^- (\varphi_{i,1})$, equal to~\eqref{eq:lim11} or~\eqref{eq:lim22}; otherwise, $\inf_{n \in\mbb{N}} \alpha_{2n+1}^- (\varphi_{i,1}) = \alpha_1^-(\varphi_{i,1})$.

 In a similar way, consider $k$ \emph{even}; then $2\floor{(k-1)/2} = k-2$, $2\ceil{(k-1)/2} = 2\floor{k/2} = 2\ceil{k/2} = k$. Thus, $\alpha_{k}^- (\varphi_{i,1})$ is, for $k$ even,
  \begin{equation}
    \frac{h_X(\varphi_{i,1}) + \frac{\abs{\lambda_i} - \abs{\lambda_i}^{k-1}}{1-\abs{\lambda_i}^2}h_{-BU}(-\varphi_{i,1}) + \frac{1-\abs{\lambda_i}^{k}}{1-\abs{\lambda_i}^2}h_{-BU}(\varphi_{i,1})  }{\frac{\abs{\lambda_i} - \abs{\lambda_i}^{k+1}}{1-\abs{\lambda_i}^2}h_{\bar{W}}(-\varphi_{i,1}) + \frac{1 - \abs{\lambda_i}^{k}}{1-\abs{\lambda_i}^2}h_{\bar{W}}(\varphi_{i,1}) },
  \end{equation}
  which has the limit~\eqref{eq:lim11} for $\abs{\lambda_i} < 1$ and
  \begin{equation}\label{eq:lim33}
   \frac{ h_{-BU}(-\varphi_{i,1}) + \abs{\lambda_i} h_{-BU}(\varphi_{i,1}) }{\abs{\lambda_i}^2 h_{\bar{W}}(-\varphi_{i,1}) + \abs{\lambda_i} h_{\bar{W}}(\varphi_{i,1})}
 \end{equation}
 for $\abs{\lambda_i} \geq 1$. If~\eqref{eq:negcond2} is met, $\inf_{n \in\mbb{N}_+} \alpha_{2n}^- (\varphi_{i,1})= \lim_{n \to\infty} \alpha_{2n}^- (\varphi_{i,1})$, equal to~\eqref{eq:lim11} or~\eqref{eq:lim33}; otherwise, $\inf_{n \in\mbb{N}_+} \alpha_{2n}^- (\varphi_{i,1}) = \alpha_2^-(\varphi_{i,1})$, which is
 \begin{equation}
   \frac{h_X(\varphi_{i,1}) + h_{-BU}(\varphi_{i,1}) }{ \abs{\lambda_i} h_{\bar{W}}(-\varphi_{i,1}) +h_{\bar{W}}(\varphi_{i,1})}.
 \end{equation}
 The corresponding limits associated with $-\varphi_{i,1}$ are obtained by performing a change of sign in the support function arguments.
 
 \hfill\IEEEQED

\subsection{Proof of Theorem~\ref{thm:comp1}}

The proof follows the same line of argument as that of Theorem~\ref{thm:main}, albeit starting from the support function of $S^\alpha_k$ in the directions $\pm\psi_{i,1}^{\ell_0+k}$ and $\pm\psi_{i,2}^{\ell_0+k}$:
\begin{multline}
(\forall j \in \{1,2\}, k \in \mbb{N})\ {h}_{S_k^{\alpha}}(\pm\psi^{\ell_0+k}_{i,j}) \leq {h}_X(\pm\psi^{\ell_0+k}_{i,j})  \\+ \sum_{\ell=0}^{k-2} {h}_{A^\ell (-BU)}(\pm\psi^{\ell_0+\ell}_{i,j}) - \sum_{\ell=0}^{k-1} {h}_{A^\ell (\alpha\bar{W})}(\pm\psi^{\ell_0+\ell}_{i,j}) ,
\end{multline}
which is the same as
\begin{multline}
(\forall j \in \{1,2\}, k \in \mbb{N})\ {h}_{S_k^{\alpha}}(\pm\psi^{\ell_0+k}_{i,j}) \leq {h}_X(\pm\psi^{\ell_0+k}_{i,j})  \\+ s_{0,k-2}(\rho_i) {h}_{-BU}(\pm\psi^{\ell_0}_{i,j}) - \alpha s_{0,k-1}(\rho_i) {h}_{\bar{W}}(\pm\psi^{\ell_0}_{i,j}) .
\end{multline}
Thus, $S_k^\alpha = \emptyset$ if there exists an $\ell_0^\prime \in \{0,\dots,M_i-1\}$ such that
\begin{align}
  {h}_{S_k^{\alpha}}(+\psi^{\ell_0^\prime+k}_{i,1}) < 0 \ &\text{and} \ {h}_{S_k^{\alpha}}(-\psi^{\ell_0^\prime+k}_{i,1}) < 0,\label{eq:condc1} \\
  \intertext{\emph{or}, there exists an $\ell_0^{\prime\prime} \in \{0,\dots,M_i-1\}$ such that}
  {h}_{S_k^{\alpha}}(+\psi^{\ell_0^{\prime\prime}+k}_{i,2}) < 0 \ &\text{and} \ {h}_{S_k^{\alpha}}(-\psi^{\ell_0^{\prime\prime}+k}_{i,2}) < 0.\label{eq:condc2}
\end{align}
Condition~\eqref{eq:condc1} is equivalent to
\ifCLASSOPTIONonecolumn
\begin{equation}
  (\exists \ell_0^\prime \in \{0,\dots,M_i-1\}) 
  \begin{dcases}
    \alpha > \alpha^{c}_{k}(+\psi_{i,1}^{\ell_0^\prime}) = \frac{h_X(+\psi_{i,1}^{\ell_0^\prime+k}) + s_{0,k-2}h_{-BU}(+\psi^{\ell_0^\prime}_{i,1})}{h_{\bar{W}}(+\psi_{i,1}^{\ell_0^\prime})},\\
    \alpha > \alpha^{c}_{k}(-\psi_{i,1}^{\ell_0^\prime}) = \frac{h_X(-\psi_{i,1}^{\ell_0^\prime+k}) + s_{0,k-2}h_{-BU}(-\psi^{\ell_0^\prime}_{i,1})}{h_{\bar{W}}(-\psi_{i,1}^{\ell_0^\prime})},
    \end{dcases}
  \end{equation}
\else
\begin{multline}
  (\exists \ell_0^\prime \in \{0,\dots,M_i-1\}) \\
  \begin{dcases}
    \alpha > \alpha^{c}_{k}(+\psi_{i,1}^{\ell_0^\prime}) = \frac{h_X(+\psi_{i,1}^{\ell_0^\prime+k}) + s_{0,k-2}h_{-BU}(+\psi^{\ell_0^\prime}_{i,1})}{h_{\bar{W}}(+\psi_{i,1}^{\ell_0^\prime})},\\
    \alpha > \alpha^{c}_{k}(-\psi_{i,1}^{\ell_0^\prime}) = \frac{h_X(-\psi_{i,1}^{\ell_0^\prime+k}) + s_{0,k-2}h_{-BU}(-\psi^{\ell_0^\prime}_{i,1})}{h_{\bar{W}}(-\psi_{i,1}^{\ell_0^\prime})},
    \end{dcases}
  \end{multline}
  \fi%
  equivalently,
  \ifCLASSOPTIONonecolumn
\begin{equation}
    (\exists \ell_0^\prime \in \{0,\dots,M_i-1\}) \ \alpha > \max\{ \alpha^{c}_{k}(+\psi_{i,1}^{\ell_0^\prime}), \alpha^{c}_{k}(-\psi_{i,1}^{\ell_0^\prime})\},
  \end{equation}
  \else
  \begin{multline}
    (\exists \ell_0^\prime \in \{0,\dots,M_i-1\}) \ \\ \alpha > \max\{ \alpha^{c}_{k}(+\psi_{i,1}^{\ell_0^\prime}), \alpha^{c}_{k}(-\psi_{i,1}^{\ell_0^\prime})\},
  \end{multline}
  \fi
  \ie
  \begin{equation}\label{eq:ac1}
\alpha > \min_{\ell_0 \in \{0,\ldots,M_i-1\}} \max\{ \alpha^{c}_{k}(+\psi_{i,1}^{\ell_0}), \alpha^{c}_{k}(-\psi_{i,1}^{\ell_0})\}.
\end{equation}
The same treatment of~\eqref{eq:condc2} yields a similar expression:
\begin{equation}\label{eq:ac2}
\alpha > \min_{\ell_0 \in \{0,\ldots,M_i-1\}} \max\{ \alpha^{c}_{k}(+\psi_{i,2}^{\ell_0}), \alpha^{c}_{k}(-\psi_{i,2}^{\ell_0})\}.
\end{equation}
Thus, $S_k^\alpha = \emptyset$ if either~\eqref{eq:ac1} or~\eqref{eq:ac2} holds: this happens if
\begin{equation}\label{eq:ac3}
\alpha > \min_{j \in \{1,2\}} \min_{\ell_0 \in \{0,\ldots,M_i-1\}} \max\{ \alpha^{c}_{k}(+\psi_{i,j}^{\ell_0}), \alpha^{c}_{k}(-\psi_{i,j}^{\ell_0})\}.
\end{equation}
Application of~\eqref{eq:psisupport} leads to the expression~\eqref{eq:hpsi}.
\hfill\IEEEQED

\subsection{Proof of Theorem~\ref{thm:comp2}}

Consider~\eqref{eq:alphac}, initially for the positive direction $\psi^{\ell_0}_{i,j}$, and let $k \to \infty$. If $\rho_i > 1$ then the second summand is, for a given $\ell_0$,
\begin{equation}
\left( \frac{1-\rho_i^{k-1}}{1-\rho_i^k} \right) \frac{h_{-BU}(\psi^{\ell_0}_{i,j})}{h_{\bar{W}}(\psi^{\ell_0}_{i,j})}  \to \frac{1}{\rho_i}\frac{h_{-BU}(\psi^{\ell_0}_{i,j})}{h_{\bar{W}}(\psi^{\ell_0}_{i,j})}. 
  \end{equation}
  For the first summand in~\eqref{eq:alphac}, we need to find
  \begin{equation}\label{eq:lim}
\lim_{k\to\infty} \left( \frac{1-\rho_i}{1-\rho_i^k} \right) \frac{h_{X}(\psi_{i,j}^{\ell_0+k})}{h_{\bar{W}}(\psi^{\ell_0}_{i,j})}.
    \end{equation}
    Note that $\lim_{k \to \infty} \frac{1-\rho_i}{1-\rho_i^k} = 0$ when $\rho_i > 1$, so to determine that the limit~\eqref{eq:lim} exists and is equal to zero, we just need to verify that
    \begin{equation}
    (\forall k \in \mbb{N}_+) \ \frac{h_{X}(\psi_{i,j}^{\ell_0+k})}{h_{\bar{W}}(\psi^{\ell_0}_{i,j})} > 0.
  \end{equation}
Since $X$ and $W$ are PC-sets (Assumption~\ref{assump:basic}), then $0 < h_X(z) < +\infty$ for all $z \in \mbb{R}^n$ and likewise $0 < h_{\bar{W}}(z) < +\infty$. Therefore, for $\rho_i > 1$,
  \begin{equation}
     \lim_{k \to \infty} {\alpha}_k^c( \psi_{i,j}^{\ell_0}) = \frac{1}{\rho_i}\frac{h_{-BU}( \psi^{\ell_0}_{i,j})}{h_{\bar{W}}( \psi^{\ell_0}_{i,j})} \eqqcolon {\alpha}_\infty^{c}(\psi^{\ell_0}_{i,j}).
  \end{equation}
  The same outcome is easily established for the $\rho_i = 1$ case. Now, for a given $\ell_0$, compare ${\alpha}_\infty^c(\psi^{\ell_0}_{i,j})$ and ${\alpha}_k^c(\psi^{\ell_0}_{i,j})$ for some $k \geq 1$:
  \ifCLASSOPTIONonecolumn
\begin{equation} 
      {\alpha}_\infty^c(\psi^{\ell_0}_{i,j}) - {\alpha}_k^c(\psi^{\ell_0}_{i,j})  = \frac{1}{\rho_i}\frac{h_{-BU}(\psi^{\ell_0}_{i,j})}{h_{\bar{W}}(\psi^{\ell_0}_{i,j})}
      - \frac{h_X(\psi^{\ell_0+k}_{i,j}) + s_{0,k-2}(\rho_i) h_{-BU}(\psi^{\ell_0}_{i,j}) }{s_{0,k-1}(\rho_i) h_{\bar{W}}(\psi^{\ell_0}_{i,j})}.
    \end{equation}
  \else
  \begin{multline} 
      {\alpha}_\infty^c(\psi^{\ell_0}_{i,j}) - {\alpha}_k^c(\psi^{\ell_0}_{i,j})  = \frac{1}{\rho_i}\frac{h_{-BU}(\psi^{\ell_0}_{i,j})}{h_{\bar{W}}(\psi^{\ell_0}_{i,j})}\\
      - \frac{h_X(\psi^{\ell_0+k}_{i,j}) + s_{0,k-2}(\rho_i) h_{-BU}(\psi^{\ell_0}_{i,j}) }{s_{0,k-1}(\rho_i) h_{\bar{W}}(\psi^{\ell_0}_{i,j})}.
    \end{multline}
    \fi
    This is negative if, and only if,
    \ifCLASSOPTIONonecolumn
    \begin{equation}
      s_{0,k-1} (\rho_i) h_{-BU}( \psi^{\ell_0}_{i,j}) - \rho_i [h_X(\psi^{\ell_0+k}_{i,j}) + s_{0,k-2}(\rho_i) h_{-BU}(\psi^{\ell_0}_{i,j})] < 0,
    \end{equation}
    \else
    \begin{multline}
      s_{0,k-1} (\rho_i) h_{-BU}( \psi^{\ell_0}_{i,j}) \\- \rho_i [h_X(\psi^{\ell_0+k}_{i,j}) + s_{0,k-2}(\rho_i) h_{-BU}(\psi^{\ell_0}_{i,j})] < 0,
    \end{multline}
    \fi
      which is equivalent to
      \begin{equation}
        \rho_i h_X(\psi^{\ell_0+k}_{i,j}) > \left( s_{0,k-1} - \rho_i s_{0,k-2} \right) h_{-BU}(\psi^{\ell_0}_{i,j}).
        \end{equation}
        But $s_{0,k-1} - \rho_is_{0,k-2} = s_{0,k-1} - s_{1,k-1} = \rho_i^0 = 1$, so an equivalent condition is
        \begin{equation}
        \rho_i h_X(\psi^{\ell_0+k}_{i,j}) > h_{-BU}(\psi^{\ell_0}_{i,j})
      \end{equation}
      in order that ${\alpha}_\infty^c(\psi^{\ell_0}_{i,j}) < {\alpha}_k^c(\psi^{\ell_0}_{i,j})$ for an arbitrary $k \in \mbb{N}_+$ and $\ell_0 \in \{0,\dots,M_i\}$. The same analysis applied to the opposite direction $-\psi^{\ell_0}_{i,j}$ says if
      \begin{equation}
        \rho_i h_X(-\psi^{\ell_0+k}_{i,j}) > h_{-BU}(-\psi^{\ell_0}_{i,j})
      \end{equation}
      then ${\alpha}_\infty^c(-\psi^{\ell_0}_{i,j}) < {\alpha}_k^c(-\psi^{\ell_0}_{i,j})$ for an arbitrary $k \in \mbb{N}_+$ and $\ell_0 \in \{0,\dots,M_i\}$. Since $k\in\mbb{N}_+$ was arbitrary and $\psi_{i,j}^{\ell_0+k} = \psi_{i,j}^{(\ell_0+k) \bmod M_i}$, if, and only if,
      \ifCLASSOPTIONonecolumn
\begin{equation}
        (\forall \ell_0 \in \{0,\dots,M_i-1\}) \ \min_{\ell \in \{0,\dots,M_i-1\}} \rho_i h_X(\psi^{\ell}_{i,j}) > h_{-BU}(\psi^{\ell_0}_{i,j})
      \end{equation}
      \else
      \begin{multline}
        (\forall \ell_0 \in \{0,\dots,M_i-1\})\\ \min_{\ell \in \{0,\dots,M_i-1\}} \rho_i h_X(\psi^{\ell}_{i,j}) > h_{-BU}(\psi^{\ell_0}_{i,j})
      \end{multline}
      \fi
      \ie
      \begin{equation}\label{eq:hXcond1}
        \min_{\ell \in \{0,\dots,M_i-1\}} \rho_i h_X(\psi^{\ell}_{i,j}) > \max_{\ell \in \{0,\dots,M_i-1\}} h_{-BU}(\psi^{\ell}_{i,j}),
      \end{equation}
      then ${\alpha}_\infty^c(+\psi^{\ell_0}_{i,j}) < {\alpha}_k^c(+\psi^{\ell_0}_{i,j})$ for all $\ell_0 \in \{0,\dots,M_i\}$. Likewise, ${\alpha}_\infty^c(-\psi^{\ell_0}_{i,j}) < {\alpha}_k^c(-\psi^{\ell_0}_{i,j})$ for all $\ell_0 \in \{0,\dots,M_i-1\}$ if, and only if,
      \begin{equation}
        \min_{\ell \in \{0,\dots,M_i-1\}} \rho_i h_X(-\psi^{\ell}_{i,j}) > \max_{\ell \in \{0,\dots,M_i-1\}} h_{-BU}(-\psi^{\ell}_{i,j}).
      \end{equation}
      Thus,
      \begin{equation}
        \inf_{k \in \mbb{N}_+} {\alpha}_k^c(\pm \psi^{\ell_0}_{i,j})  = {\alpha}_\infty^c(\pm \psi^{\ell_0}_{i,j})
      \end{equation}
      and
      \ifCLASSOPTIONonecolumn
      \begin{equation}
        \inf_{k \in \mbb{N}_+} \bar{\alpha}_k^c(\lambda_i) =  \min_{j\in \{1,2\}} \min_{\ell_0 \in \{0,\dots,M_i-1\}} \max\{ \alpha_{\infty}^c(+\psi^{\ell_0}_{i,j}), \alpha_{\infty}^c(-\psi^{\ell_0}_{i,j})\}.
      \end{equation}
      \else
      \begin{multline}
        \inf_{k \in \mbb{N}_+} \bar{\alpha}_k^c(\lambda_i) = \\ \min_{j\in \{1,2\}} \min_{\ell_0 \in \{0,\dots,M_i-1\}} \max\{ \alpha_{\infty}^c(+\psi^{\ell_0}_{i,j}), \alpha_{\infty}^c(-\psi^{\ell_0}_{i,j})\}.
      \end{multline}
      \fi

        The case of $0 < \rho_i < 1$ yields to a different analysis. Then $\rho_i^k \to 0$, but $h_X(\psi^k_{i,j})$ in the first summand in~\eqref{eq:alphac} may not have a limit because $\psi^k_{i,j}$ does not have a limit. Nonetheless, considering the periodicity of $\psi^k_{i,j}$, let, for $j \in \{1,2\}$, $\ell_0, \ell \in \{0,\dots,M_i-1\}$
        \begin{equation}
          {\alpha}_\infty^{c, \ell}(\pm\psi^{\ell_0}_{i,j}) \coloneqq (1-\rho_i)\frac{h_{X}(\pm \psi_{i,j}^{\ell})}{h_{\bar{W}}(\pm\psi^{\ell_0}_{i,j})} + \frac{h_{-BU}(\pm\psi^{\ell_0}_{i,j})}{h_{\bar{W}}(\pm\psi^{\ell_0}_{i,j})},
  \end{equation}
such that $\{ {\alpha}_\infty^{c, 0}(+\psi^{\ell_0}_{i,j}), \dots, {\alpha}_\infty^{c, M_i-1}(+\psi^{\ell_0}_{i,j}) \}$ collects, for each $\ell_0 \in \{0,\dots,M_i-1\}$, the $M_i$ different limit values of ${\alpha}_k^c(+\psi^{\ell_0}_{i,j})$ as $k \to \infty$, and similarly in the opposite directions.

  We then seek to compare, for a given $\ell_0 \in\{0,\dots,M_i\}$ and an arbitrary $k \in \mbb{N}_+$, the values of  ${\alpha}_\infty^{c,\ell}(+\psi^{\ell_0}_{i,j})$ and ${\alpha}_k^c(+\psi^{\ell_0}_{i,j})$ (positive direction first) where 
  \begin{equation}\label{eq:lkmod}
    \ell = (\ell_0 + k) \bmod M_i.
    \end{equation}
    That way, since
    \begin{equation}\label{eq:hperiod}
      h_X(\psi^{\ell_0+k}_{i,j}) = h_X(\psi^{(\ell_0+k)\bmod M_i}_{i,j}) = h_X(\psi^{\ell}_{i,j}),
    \end{equation}
    both ${\alpha}_\infty^{c,\ell}(+\psi^{\ell_0}_{i,j})$ and
    ${\alpha}_k^c(+\psi^{\ell_0}_{i,j})$ involve evaluations of $h_X$ in the same direction. The value of ${\alpha}_\infty^{c,\ell}(+\psi^{\ell_0}_{i,j})$ is strictly less than ${\alpha}_k^c(+\psi^{\ell_0}_{i,j})$ if, and only if,
    \ifCLASSOPTIONonecolumn
    \begin{equation}
    h_X(\psi^{\ell_0+k}_{i,j}) > (1-\rho_i)s_{0,k-1} h_X(\psi_{i,j}^{\ell})  + (s_{0,k-1} - s_{0,k-2})h_{-BU}(\psi_{i,j}^{\ell_0}).
  \end{equation}
    \else
  \begin{multline}
    h_X(\psi^{\ell_0+k}_{i,j}) > (1-\rho_i)s_{0,k-1} h_X(\psi_{i,j}^{\ell}) \\ + (s_{0,k-1} - s_{0,k-2})h_{-BU}(\psi_{i,j}^{\ell_0}).
  \end{multline}
  \fi%
Note that $(1-\rho_i)s_{0,k-1} = s_{0,k-1} - s_{1,k}  = 1 - \rho_i^k$, and $s_{0,k-1} - s_{0,k-2} = \rho_i^{k-1}$. Hence, this inequality is
  \begin{equation}\label{eq:newcond2}
    h_X(+\psi^{\ell_0+k}_{i,j}) > (1-\rho_i^k) h_X(+\psi_{i,j}^{\ell})  + \rho_i^{k-1} h_{-BU}(+\psi_{i,j}^{\ell_0}),
  \end{equation}
but, because of~\eqref{eq:hperiod}, inequality~\eqref{eq:newcond2} reduces to
    \begin{equation}\label{eq:newcond}
    0 > \rho_i^{k-1}\left( h_{-BU}(+\psi_{i,j}^{\ell_0}) - \rho_i h_X(+\psi_{i,j}^{(\ell_0+k)\bmod M_i})\right).
  \end{equation}
  Since $\rho_i\in (0,1)$, $\rho^{k-1} > 0$  and, therefore,
  \begin{equation}
  {\alpha}_\infty^{c,(\ell_0 + k)\bmod M_i}(+\psi^{\ell_0}_{i,j}) < {\alpha}_k^c(+\psi^{\ell_0}_{i,j})%
\end{equation}%
for a given $j\in\{1,2\}, \ell_0\in\{0,\dots,M_i-1\}$ and $k \in \mbb{N}_+$, if, and only if,
  \begin{equation}
    \rho_i h_X(\psi_{i,j}^{(\ell_0+k)\bmod M_i}) > h_{-BU}(\varphi_{i,j}^{\ell_0}).
  \end{equation}
  In fact,
  \begin{equation}\label{eq:alphainfdec}
  {\alpha}_\infty^{c,(\ell_0 + k)\bmod M_i}(+\psi^{\ell_0}_{i,j}) < {\alpha}_k^c(+\psi^{\ell_0}_{i,j})%
\end{equation}%
for \emph{all} $j\in\{1,2\}, \ell_0\in\{0,\dots,M_i-1\}$ and $k \in \mbb{N}_+$, if
  \begin{equation}\label{eq:hXcond2}
    \min_{\ell \in \{0,\dots,M_i-1\}} \rho_i h_X(\psi_{i,j}^{\ell}) > \max_{\ell \in \{0,\dots,M_i-1\}} h_{-BU}(\varphi_{i,j}^{\ell}),
\end{equation}
which is condition~\eqref{eq:hXcond1} once again.

Comparison between ${\alpha}_\infty^{c,\ell}(-\psi^{\ell_0}_{i,j})$
and ${\alpha}_k^c(-\psi^{\ell_0}_{i,j})$ yields the same result and
condition with the signs of $\psi$ reversed.

The result~\eqref{eq:alphainfdec} means that, for a given $\ell_0 \in \{0,\ldots,M_i-1\}$,
\begin{align}
  {\alpha}_\infty^{c,(\ell_0 + 1)\bmod M_i}(+\psi^{\ell_0}_{i,j}) &< {\alpha}_1^c(+\psi^{\ell_0}_{i,j}) \\
  {\alpha}_\infty^{c,(\ell_0 + 2)\bmod M_i}(+\psi^{\ell_0}_{i,j}) &< {\alpha}_2^c(+\psi^{\ell_0}_{i,j})\\
                                                                  &\vdots \notag\\
  {\alpha}_\infty^{c,(\ell_0 + M_i)\bmod M_i}(+\psi^{\ell_0}_{i,j}) &< {\alpha}_{M_i}^c(+\psi^{\ell_0}_{i,j})\\
  {\alpha}_\infty^{c,(\ell_0 + 1)\bmod M_i}(+\psi^{\ell_0}_{i,j}) &< {\alpha}_{1+M_i}^c(+\psi^{\ell_0}_{i,j})\\
  {\alpha}_\infty^{c,(\ell_0 + 2)\bmod M_i}(+\psi^{\ell_0}_{i,j}) &< {\alpha}_{2+M_i}^c(+\psi^{\ell_0}_{i,j}),
  \end{align}
  and so on, so that, in general,
  \begin{equation}
    {\alpha}_\infty^{c,(\ell_0 + k_0)\bmod M_i}(+\psi^{\ell_0}_{i,j}) < {\alpha}_{k_0+n M_i}^c(+\psi^{\ell_0}_{i,j})
  \end{equation}
  for all $n \in \mbb{N}$ and where $k_0 \in \{1,\dots,M_i\}$. Thus, if~\eqref{eq:hXcond2} holds, then for all $k_0 \in \{1,\dots,M_i\}$ and $\ell_0 \in \{0,\dots,M_i-1\}$,%
      \begin{equation}
        \inf_{n \in \mbb{N}} {\alpha}_{k_0+nM_i} ^c(+\psi^{\ell_0}_{i,j})  = {\alpha}_\infty^{c,(\ell_0+k_0)\bmod M_i}(+\psi^{\ell_0}_{i,j})
      \end{equation}
      and similarly for $-\psi_{i,j}$. Therefore, when $\rho_i < 1$,
      \ifCLASSOPTIONonecolumn
\begin{equation}
        \inf_{k\in\mbb{N}_+} \bar{\alpha}_k^c(\lambda_i)) =  \min_{j \in \{1,2\}} \min_{\ell_0, \ell \in \{0,\dots,M_i-1\}}  \max\{ {\alpha}_\infty^{c,\ell}(+\psi^{\ell_0}_{i,j}), {\alpha}_\infty^{c,\ell}(-\psi^{\ell_0}_{i,j}) \}.
      \end{equation}
      \else
      \begin{multline}
        \inf_{k\in\mbb{N}_+} \bar{\alpha}_k^c(\lambda_i)) = \\ \min_{j \in \{1,2\}} \min_{\ell_0, \ell \in \{0,\dots,M_i-1\}}  \max\{ {\alpha}_\infty^{c,\ell}(+\psi^{\ell_0}_{i,j}), {\alpha}_\infty^{c,\ell}(-\psi^{\ell_0}_{i,j}) \}.
      \end{multline}
      \fi
\hfill\IEEEQED

\bibliographystyle{IEEEtran}
\bibliography{extracted.bib}


\begin{IEEEbiography}[{\includegraphics
[width=1in,height=1.25in,clip,
keepaspectratio]{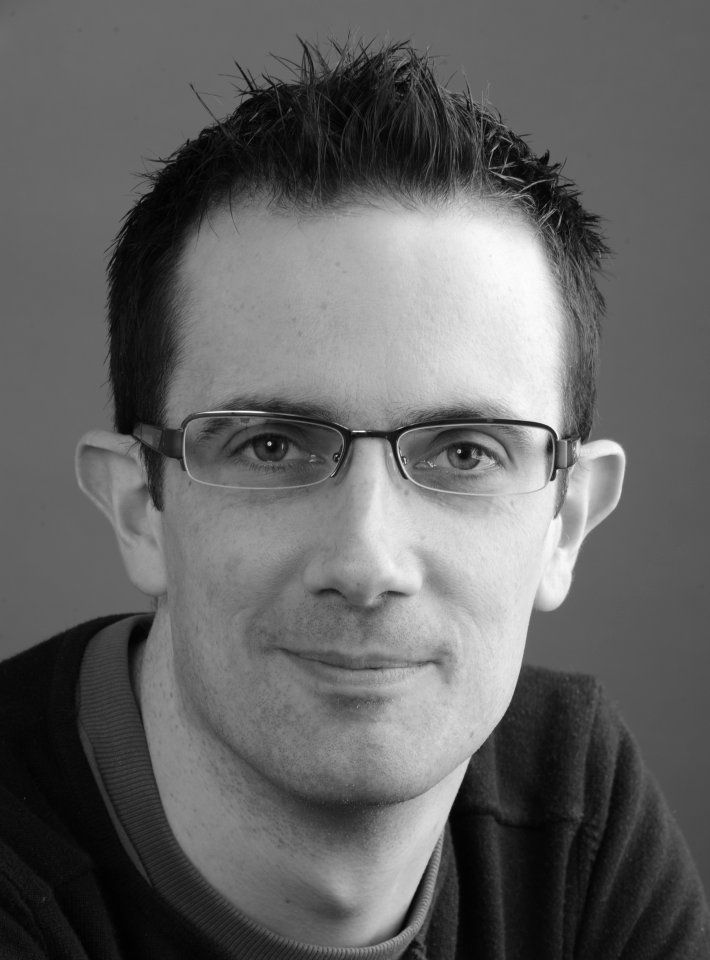}}]
{Paul Trodden} received the M.Eng. degree in Engineering Science from the University of Oxford in 2003 and the Ph.D. degree in Aerospace Engineering from the University of Bristol in 2009. He subsequently held a PDRA positions in the Department of Aerospace Engineering, University of Bristol (2009--10) and the School of Mathematics, University of Edinburgh (2010--12). Since 2012, he has been a Lecturer and then Senior Lecturer with the School of Electrical and Electronic Engineering, University of Sheffield. His research interests include model predictive and optimization-based control, especially distributed and robust forms, and applications of control and optimization to aerospace and power and energy systems. He is a member of the IEEE Control Systems Society Conference Editorial Board and an Associate Editor for \emph{IET Smart Grid}.
\end{IEEEbiography}


\begin{IEEEbiography}[{\includegraphics
[width=1in,height=1.25in,clip,
keepaspectratio]{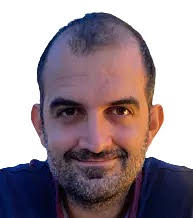}}]
{Jos\'e M. Maestre}
holds a PhD from the University of Seville, where he currently serves as a full professor. He has held positions at TU Delft, the University of Pavia, Kyoto University, and the Tokyo Institute of Technology. He is the author of \emph{Service Robotics within the Digital Home} (Springer, 2011), \emph{A Programar se Aprende Jugando} (Paraninfo, 2017), \emph{Sistemas de Medida y Regulaci\'on} (Paraninfo, 2018), and \emph{Model Predictive Control} (Springer, 2025). He is also the editor of \emph{Distributed Model Predictive Control Made Easy} (Springer, 2014) and \emph{Control Systems Benchmarks} (Springer, 2025).
 
His research focuses on the control of distributed cyber-physical systems, with a special emphasis on integrating heterogeneous agents into the control loop. He has published more than 200 journal and conference papers and has led multiple research projects. His achievements have been recognized with several awards and honors, including the Spanish Royal Academy of Engineering medal for his contributions to predictive control in large-scale systems and the distinction of becoming the youngest full professor in the Spanish university system in 2020.
\end{IEEEbiography}


\begin{IEEEbiography}[{\includegraphics
[width=1in,height=1.25in,clip,
keepaspectratio]{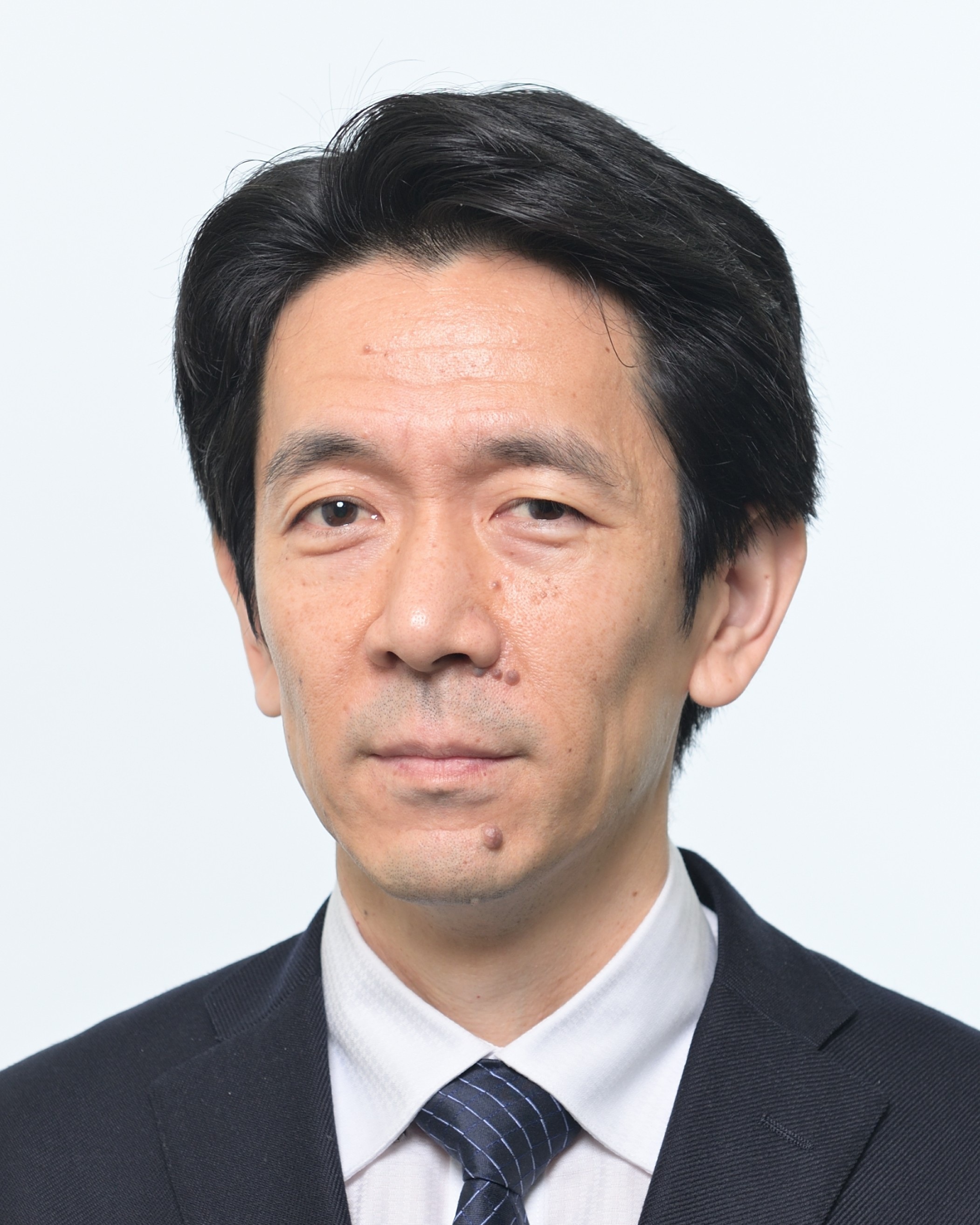}}]
{Hideaki Ishii} received the
M.Eng. degree from Kyoto University, Japan, in 1998, 
and the Ph.D. degree from the University of Toronto, Canada, in 2002. 
He was a Postdoctoral Research
Associate at the University of Illinois at Urbana-Champaign, USA, in 2001-2004, and a Research Associate at The University of Tokyo, Japan,
in 2004-2007.
He was an Associate Professor and then a Professor 
at the Tokyo Institute of Technology, Japan, in 2007--2024.
Currently, he is a Professor at the Department of
Information Physics and Computing, The University of Tokyo, Japan.
He was a Humboldt Research Fellow at the University of Stuttgart
in 2014--2015. He has also held visiting positions at CNR-IEIIT at
the Politecnico di Torino, the Technical University of Berlin, and
the City University of Hong Kong. His research interests
include networked control systems, multiagent systems, distributed algorithms,
and cyber-security of control systems.

Dr. Ishii has served as an Associate Editor for \emph{Automatica}, 
the \emph{IEEE Control Systems Letters}, the \emph{IEEE Transactions on Automatic Control}, 
the \emph{IEEE Transactions on Control of Network Systems},
and the \emph{Mathematics of Control, Signals, and Systems}.
He was a Vice President for the IEEE Control Systems Society (CSS) in 2022--2023, the Chair of the IFAC Coordinating Committee on Systems and Signals in 2017--2023, and the Chair of the IFAC Technical Committee
on Networked Systems for 2011--2017.
He served as the IPC Chair for the IFAC World Congress 2023 held in Yokohama, Japan.
He received the IEEE Control Systems Magazine Outstanding Paper
Award in 2015. Dr. Ishii is an IEEE Fellow. 
\end{IEEEbiography}

\vfill

\end{document}